\documentclass[12pt]{article}

\usepackage{graphicx,psfrag,epsf,color}
\usepackage{amsmath,amssymb,amsfonts}
\usepackage{array}
\usepackage{cite}
\usepackage{multirow}
\usepackage{rotating}
\usepackage{url} % in oder to include the url of the SLAC group

\newcommand{\grtsim}{\mbox{\raisebox{-3pt}{$\stackrel{>}{\sim}$}}}
\newcommand{\lessim}{\mbox{\raisebox{-3pt}{$\stackrel{<}{\sim}$}}}

\begin{document}

\allowdisplaybreaks

\begin{titlepage}

\begin{flushright}
{\small
TUM-HEP-955/14\\
TTK-14-22\\
SFB/CPP-14-70\\
IFIC/14-59\\[0.2cm]

November 22, 2014
}
\end{flushright}

\vskip1cm
\begin{center}
\Large\bf\boldmath
Heavy neutralino relic abundance with Sommerfeld enhancements -- 
a study of pMSSM scenarios 
\end{center}

\vspace{0.8cm}
\begin{center}
{\sc M.~Beneke$^{a,b}$, C.~Hellmann$^{a,b}$} and 
{\sc P. Ruiz-Femen\'\i a$^{c}$}\\[5mm]
{\it ${}^a$Physik Department T31,\\
James-Franck-Stra\ss e, 
Technische Universit\"at M\"unchen,\\
D--85748 Garching, Germany\\
\vspace{0.3cm}
${}^b$Institut f\"ur Theoretische Teilchenphysik und 
Kosmologie,\\
RWTH Aachen University, D--52056 Aachen, Germany\\
\vspace{0.3cm}
${}^c$Instituto de F\'\i sica Corpuscular (IFIC), 
CSIC-Universitat de Val\`encia \\
Apdo. Correos 22085, E-46071 Valencia, Spain}\\[0.3cm]
\end{center}

\vspace{1cm}
\begin{abstract}
\vskip0.2cm\noindent
We present a detailed discussion of Sommerfeld enhancements in neutralino
dark matter relic abundance calculations for several popular benchmark
scenarios in the general MSSM. Our analysis is focused on models
with heavy wino- and higgsino-like neutralino LSP and models interpolating
between these two scenarios. This work is the first phenomenological 
application of effective field theory
methods that we have developed in earlier work and that allow for the
consistent study of Sommerfeld enhancements in non-relativistic neutralino and
chargino co-annihilation reactions within the general MSSM, away from 
the pure-wino and pure-higgsino limits.
\end{abstract}
\end{titlepage}

%\setcounter{page}{1}

%%%%%%%%%%%%%%%%%%%%%%%%%%%%%%%%%%%%%%%%%%%%%%%%%%%%%%%%%%%%%%%%%%%%%%%%%%%%%%

%\tableofcontents

\section{Introduction}
\label{sec:introduction}

The experimental determination of the cold dark matter density in our Universe
has reached percent level accuracy, $\Omega_\text{cdm} h^2 =
0.1187\pm 0.0017$, taking recent PLANCK data into account \cite{Ade:2013zuv}.
While the nature and origin of the cosmic dark matter component are still
unknown it is intriguing that the observed abundance can be explained rather
naturally as thermal relic of a TeV scale particle with weak
interaction strength.
A central ingredient in the relic abundance calculation of a particle dark
matter (DM) candidate is its pair-annihilation rate. Consequently the increasing
experimental precision on $\Omega_\text{cdm} h^2$ has triggered a particular
interest in the calculation of radiative corrections to particle DM
annihilation cross sections.

Probably the best motivated and certainly one of the most studied particle dark 
matter candidates is the neutralino LSP $(\chi^0_1)$ in the minimal
supersymmetric standard model (MSSM) \cite{Jungman:1995df, Bertone:2004pz}.
Several codes \cite{Gondolo:2004sc ,Belanger:2010gh} allow for the calculation
of the $\chi^0_1$ relic abundance in the general MSSM, currently relying on
co-annihilation rates calculated at tree-level.
The calculation of radiative corrections to these rates
follows two different directions.
On the one hand, a great effort is undertaken in the calculation of
next-to-leading order co-annihilation rates in fixed order perturbation theory
in the MSSM.
The complete next-to-leading order SUSY QCD corrections in $\chi^0_1$
co-annihilations with potentially nearly mass degenerate charginos and
sfermions has been performed
\cite{Herrmann:2007ku, Herrmann:2009wk, Herrmann:2009mp, Harz:2012fz,
Herrmann:2014kma} and the first steps in the calculation of the full one-loop
electroweak corrections are undertaken
\cite{Baro:2007em, Baro:2008em, Baro:2009na }.
On the other hand it has been noted some time ago that in non-relativistic
dark matter pair-annihilations a certain class of radiative corrections can be
enhanced and requires a systematic resummation up to all loop-orders
\cite{Hisano:2004ds, Hisano:2006nn}.
The resulting Sommerfeld enhancement arises naturally in
theories with light mediator exchange between the co-annihilating
non-relativistic dark matter particles prior to their actual annihilation
reaction.
For heavy $\chi^0_1$ dark matter this effect should be addressed generically
in the relic abundance calculation as well as in indirect detection rates:
in both cases the annihilating particles move at non-relativistic velocities
and the mutual exchange of electroweak gauge bosons -- and to a 
lesser extent, Higgs bosons --  prior to the actual
annihilation gives rise to long-range potential interactions eventually
requiring a systematic resummation of certain contributions up to all loop
orders.
In context of $\chi^0_1$ pair annihilation reactions, Sommerfeld enhancements
have been first addressed in the pure-wino and pure-higgsino $\chi^0_1$
scenarios in \cite{Hisano:2004ds, Hisano:2006nn} and were subsequently studied
in \cite{Hryczuk:2010zi, Hryczuk:2011tq}.
The particular relevance in $\chi^0_1$ indirect detection has
been investigated for the pure-wino case in
\cite{Cohen:2013ama, Fan:2013faa, Hryczuk:2014hpa}.

In \cite{Beneke:2012tg, Hellmann:2013jxa, paperIII} we have developed a
formalism that allows to systematically address the calculation of 
enhanced radiative
corrections in non-relativistic neutralino/chargino pair-annihi\-la\-ti\-on
reactions in the general MSSM by means of a non-relativistic effective field
theory approach, where our particular focus is the consistent calculation of
Sommerfeld enhancements. By ``general MSSM'' we imply that the lightest 
neutralino can be an arbitrary admixture of wino, higgsino and bino.  
Analytic results for the short-distance coefficients encoding hard
tree-level annihilation reactions of non-relativistic co-annihilating
neutralinos and charginos including $P$- and next-to-next-to-leading order
$S$-wave rates are given in \cite{Beneke:2012tg, Hellmann:2013jxa}.
Corresponding analytic expressions for the long-range potential interactions
eventually causing enhanced annihilation rates are presented 
in~\cite{paperIII}. In the latter work we also describe the technical 
details involved in a precise
determination of Sommerfeld enhancements in the $\chi^0_1$ relic abundance
calculation. It is worth noting that in addition to covering the general 
case of $\chi^0_1$ being an arbitrary admixture of the electroweak 
gaugino eigenstates, our approach extends previous investigations on 
the subject in several other aspects, such as the
consistent treatment of off-diagonal annihilation rates, the separation 
into $S$- and $P$-wave components with their own, separate Sommerfeld 
factors, and the ability to deal with many nearly mass-degenerate states.

The purpose of this paper is a detailed investigation and discussion of
Sommerfeld enhancements in the $\chi^0_1$ relic abundance calculation in some
popular MSSM scenarios.
The underlying physics effects are analysed in detail in each step of the
calculation.
This allows to illustrate the general use of our method
\cite{Beneke:2012tg, Hellmann:2013jxa, paperIII}
applicable in the general MSSM and to address the question of viability of
popular MSSM scenarios in light of a consistent treatment of the Sommerfeld
effect.
We choose to consider three scenarios taken from the set of
Snowmass pMSSM benchmark models \cite{Cahill-Rowley:2013gca}.
These models pass all constraints from so far unsuccessful SUSY searches at
the LHC, additional collider, flavour and precision measurement bounds as
well as constraints from dark matter direct detection experiments and indirect
searches. The neutralino LSP relic abundance within these models, calculated
from perturbative annihilation rates, is not larger than the WMAP bound, but
can be smaller than the experimentally measured value. The latter allows for
the case that neutralino dark matter does not make up all the cosmic cold dark
matter.
In addition to these benchmark scenarios we investigate the Sommerfeld
enhancements in neutralino/chargino co-annihilations in a set of
models interpolating between a scenario with almost pure-higgsino $\chi^0_1$
to a wino-like $\chi^0_1$ model. The MSSM spectra for the models on this
``higgsino-to-wino'' trajectory are generated with 
DarkSUSY \cite{Gondolo:2004sc}.
As our work allows for the first time a consistent study of the Sommerfeld
effect on the relic abundance calculation for models with mixed wino-higgsino
neutralino LSP we provide an extensive discussion of the Sommerfeld effect in
such a scenario.

Throughout this work we neither include thermal effects nor the effect of
running couplings. As regards thermal effects in context of dark matter relic
abundance calculations including Sommerfeld enhancements, the temperature
dependence of the gauge boson masses has, for instance, been considered in
\cite{Cirelli:2007xd, Hryczuk:2010zi}. Concerning the running of couplings,
this effect can in principle be relevant to Sommerfeld-enhanced
rates as well: the annihilation process involves the mass scale of
the co-annihilating particles, that is associated with the hard annihilation
reaction, as well as the much smaller scale of the non-relativistic kinetic 
energies of the co-annihilating particle pairs and the masses of the 
exchanged particles. The latter scales are connected with the physics that
causes the Sommerfeld enhancement.
Both the effect from thermal corrections and from running couplings will be
investigated in future work.

The structure of this paper is as follows. In Sec.~\ref{sec:res_wino} we
consider the case of a wino-like $\chi^0_1$ benchmark scenario taken
from \cite{Cahill-Rowley:2013gca}, followed by the investigation of a
higgsino-like $\chi^0_1$ benchmark spectrum in Sec.~\ref{sec:res_higgsino}.
In both cases we compare to results obtained in the well-studied
``pure'' wino and higgsino scenarios where the $\chi^0_1$ is assumed to be part
of an unbroken $SU(2)_L$ triplet or two unbroken $SU(2)_L$ doublets. As
Sommerfeld enhancements have been studied extensively in the particular case of
a pure wino $\chi^0_1$ in  the literature, we address the question of the
validity of conclusions inferred from these pure wino and higgsino scenarios
to wino- and higgsino-like $\chi^0_1$ spectra in the general MSSM.
In Sec.~\ref{sec:res_bino} the effect of Sommerfeld enhancements in 
co-annihilations of wino-like neutralino and chargino states in a bino-like
$\chi^0_1$ benchmark scenario is considered.
A ``higgsino-to-wino'' trajectory is defined in Sec.~\ref{sec:res_trajectory},
by introducing $13$ models that interpolate between a higgsino- and wino-like
$\chi^0_1$ spectrum while the relic density calculated from perturbative rates
is kept fixed. Our discussion here is focused on the spectra and
the obtained relic abundances omitting particular details on the Sommerfeld
enhanced co-annihilation cross sections.
The specific features of the Sommerfeld effect for a mixed wino-higgsino
$\chi^0_1$ are subsequently studied in detail in
Sec.~\ref{sec:res_winohiggsino}, where the selected spectrum is
one of the trajectory models of the preceding section.
We draw our conclusions and give an outlook to future work in
Sec.~\ref{sec:summary}.

The present paper is intentionally stripped of all technical details 
underlying the computation of the results and focuses on 
the nature of the Sommerfeld enhancement and its physics interpretation. 
The reader interested in technical aspects is referred to 
\cite{Beneke:2012tg, Hellmann:2013jxa} for the computation of the 
annihilation cross sections and in particular to \cite{paperIII} for 
the computation of the Sommerfeld enhancements and the solution of 
the multi-channel Schr\"odinger equation.

%-----------------------------------------------------------------------------
\section{Wino-like $\chi^0_1$}
\label{sec:res_wino}

Wino-like $\chi^0_1$ dark matter arranges into an approximate $SU(2)_L$ fermion
triplet together with the two chargino states $\chi^\pm_1$. In the
$SU(2)_L\times U(1)_Y$ symmetric limit the triplet would be assigned zero
hypercharge. All states $\chi^0_1,\chi^\pm_1$ share the same
$\mathcal O($TeV$)$ mass scale, characterised by the wino mass parameter $M_2$,
$m_\chi\sim\vert M_2\vert$. Electroweak symmetry-breaking introduces a
small mass splitting between the neutral and the charged components of the
triplet. The tree-level mass splitting happens to be very small, 
$\mathcal O(m_W^4/m^3_\text{SUSY})$, and the one-loop radiative corrections
dominate over the tree-level splitting.

A pMSSM scenario with wino-like $\chi^0_1$ is provided by the SUSY spectrum with
model ID $2392587$ in \cite{Cahill-Rowley:2013gca}. A measure for the 
wino fraction of a given neutralino LSP state is the square of
the modulus of the neutralino mixing-matrix entry $Z_{N\,21}$. For pMSSM scenario
$2392587$ the $\chi^0_1$ constitutes a rather pure wino,
$\vert Z_{N\,21}\vert^2 = 0.999$, with a mass $m_{\rm LSP}\equiv 
m_{\chi^0_1} = 1650.664\,$GeV.
The mass of the chargino partner $\chi^\pm_1$ is given by
$m_{\chi^+_1} = 1650.819\,$GeV, such that $\delta m  = m_{\chi^+_1} - m_{\chi^0_1}$
turns out to be $0.155\,$GeV.
Without any modification these values are taken from the spectrum card provided
by \cite{Cahill-Rowley:2013gca} where the mass parameters
refer to the $\overline{\text{DR}}$-scheme.
As the precise sub $\mathcal O($GeV$)$-scale $\chi^0_1\chi^\pm_1$ mass splitting 
is an essential ingredient in the calculation of the Sommerfeld-enhanced
co-annihilation rates we have to assume an accuracy of the given
mass spectrum at the level of $10\,$MeV for our analysis of the Sommerfeld
enhancement in the pMSSM scenario to be meaningful.
A rigorous analysis of Sommerfeld-enhanced
co-annihilation processes in a given model should refer to the on-shell
mass spectrum of the neutralino and chargino states instead of
$\overline{\text{DR}}$-parameters, where a sub-GeV scale precision of the
mass parameters requires the consideration of one-loop
renormalised quantities. For reference purposes, however, we do not modify the
publicly available $\overline{\text{DR}}$-spectra of
\cite{Cahill-Rowley:2013gca} for all three pMSSM models discussed here.

In the context of minimal dark matter models \cite{Cirelli:2007xd}, wino dark
matter is realised as the neutral component of an approximate 
$SU(2)_L$ triplet state as well. In contrast to MSSM 
scenarios with wino-like $\chi^0_1$, the
$SU(2)_L$ triplet minimal dark matter models (referred to as ``pure-wino''
models in the following) consider interactions of the dark
matter states with the electroweak gauge bosons only.
Two-particle final states in minimal dark matter 
pair-annihilation reactions are
hence given by pairs of SM particles and the SM Higgs boson and all heavier
states above the minimal dark matter mass scale are treated as completely
decoupled.
Such a scenario agrees with the decoupling limit in a MSSM scenario with
wino-like $\chi^0_1$ LSP. To the contrary, the wino-like pMSSM model that we
consider here features non-decoupled sfermion states at the $2-3\,$TeV scale
with non-vanishing couplings of the $\chi^0_1$ and $\chi^\pm_1$ to 
sfermions and to the (heavier) Higgs states, though the latter are 
suppressed with respect to the couplings to the gauge bosons,
because any Higgs-$\chi\chi$ (tree-level) interaction takes place between the
gaugino-component of the one and the higgsino-component of the other $\chi$.
As the higgsino-like neutralino and chargino states in the pMSSM model under
consideration reside at the $\mathcal O( 3.9\,$TeV$)$ scale any\
Higgs-$\chi\chi$ interaction plays a sub-dominant role in our analysis of
pair-annihilation reactions of the wino-like $\chi^0_1$ and $\chi^\pm_1$ 
states. Due to the non-decoupled sfermion states though, some 
annihilation rates in the wino-like $\chi^0_1$ pMSSM scenario are 
reduced with respect to the pure-wino dark matter case. 

In the calculation of the relic abundance we have to take into account all
possible two-particle co-annihilation reactions between the (approximate)
$SU(2)_L$ triplet states $\chi^0_1, \chi^\pm_1$.
In addition, in the pMSSM model $2392587$, the bino-like $\chi^0_2$ is
only about $8\%$ heavier than the $\chi^0_1$, $m_{\chi^0_2} = 1781.37\,$GeV.
Hence the $\chi^0_2$ is a potentially relevant co-annihilating particle  as 
well. It turns out though, that this state eventually plays no 
role for the relic abundance, as the corresponding cross sections
are strongly suppressed with respect to those of the
wino-like particles $\chi^0_1$ and $\chi^\pm_1$ due to the much
weaker couplings of the bino-like $\chi^0$ to gauge bosons and to the remaining
$\chi^0/\chi^\pm$ states. All remaining heavier particles 
in the pMSSM scenario lie above the $2\,$TeV scale, so they are
already Boltzmann suppressed and hence practically irrelevant 
during the $\chi^0_1$ freeze-out.

Sommerfeld enhancements on the co-annihilation rates are taken into 
account by including in the multi-state Schr\"odinger equation  all 
$\chi\chi$ two-particle states with mass smaller than $M_{\rm max} = 
2\,m_{\chi^0_1} + m_{\chi^0_1} v_\text{max}^2$, where we set
$v_\text{max} = 1/3$. This choice is motivated by the fact that 
$v_\text{max}$ roughly corresponds to the $\chi^0_1$'s
mean velocity around freeze-out, hence these states are potentially 
relevant for co-annihilation processes, and can still be produced 
on-shell in a $\chi^0_1\chi^0_1$ scattering process. 
The remaining heavier two-particle states with mass above $M_{\rm max}$ 
are included in the computation of the Sommerfeld enhancement of 
the lighter states in the last loop before the annihilation, following
the method developed and discussed in \cite{paperIII}.
The $\chi\chi$-channels, whose long-distance interactions are 
treated exactly, can be classified according to their total electric charge.
The sector of neutral two-particle states comprises the 
$\chi^0_1\chi^0_1$ and $\chi^+_1\chi^-_1$
channels. In the pMSSM scenario considered here, this sector contains in addition the
$\chi^0_1\chi^0_2$ state.
In the single-charged and the double-charged sectors of a pure-wino
dark matter scenario there is only one state present in each sector,
$\chi^0_1\chi^+_1$ ($\chi^0_1\chi^-_1$) and $\chi^+_1\chi^+_1$ ($\chi^-_1\chi^-_1$), whereas in the pMSSM scenario we have to add in addition 
a second state with $\chi^0_1$ replaced by $\chi^0_2$, in agreement
with the rule above that defines the channels which enter the 
Schr\"odinger equation. Since the bino-like neutralino essentially neither 
couples to the wino-like particles nor to gauge bosons, and because 
sfermion states are rather
heavy, potential interactions as well as tree-level annihilation reactions
involving the bino-like $\chi^0_2$ are strongly suppressed with respect to the
corresponding interactions with wino-like particles $\chi^0_1,\chi^\pm_1$.
As a consequence, 
$\chi^0_2$ plays essentially no role for Sommerfeld enhancements,
and we focus the discussion that follows 
on the channels built from the wino-like $\chi^0_1$ and
$\chi^\pm_1$ states only.

In each of the  charge sectors long-range interactions due to
potential exchange of electroweak gauge bosons, photons and light Higgses are
present.\footnote{ Potentials from Higgs exchange are negligible compared to
the leading contributions from gauge bosons in the pMSSM scenario with wino-like
$\chi^0_1$, again because in any Higgs-$\chi\chi$ vertex the gaugino component
of one $\chi$ is coupled to the higgsino component of the other $\chi$.
In the wino-like $\chi^0_1$ Snowmass model the lowest-lying $\chi$'s relevant
for the Sommerfeld effect are rather pure wino-like $\chi^0$ and $\chi^\pm$
states with a very small higgsino component.}
Potential $W$-boson exchange leads to a Yukawa potential interaction
that induces transitions between the $\chi^0_1\chi^0_1$ and the $\chi^+_1\chi^-_1$
state in the neutral sector.
Hence the part of the neutral sector consisting of the channels
$\chi^0_1\chi^0_1$ and $\chi^+_1\chi^-_1$ is 
characterised by a potential matrix with non-vanishing off-diagonals which are
of the same strength as the diagonal entries. As the incoming $\chi^0_1\chi^0_1$
pair cannot build a ${}^3S_1$ or ${}^1P_1$ state, potential
interactions are responsible for transitions between the two neutral states
$\chi^0_1\chi^0_1$ and $\chi^+_1\chi^-_1$ in a ${}^1S_0$ or ${}^3P_{\cal J}$
configuration.

%---------------------------------------------------------------------------
\begin{figure}[t]
\begin{center}
\includegraphics[width=0.75\textwidth]{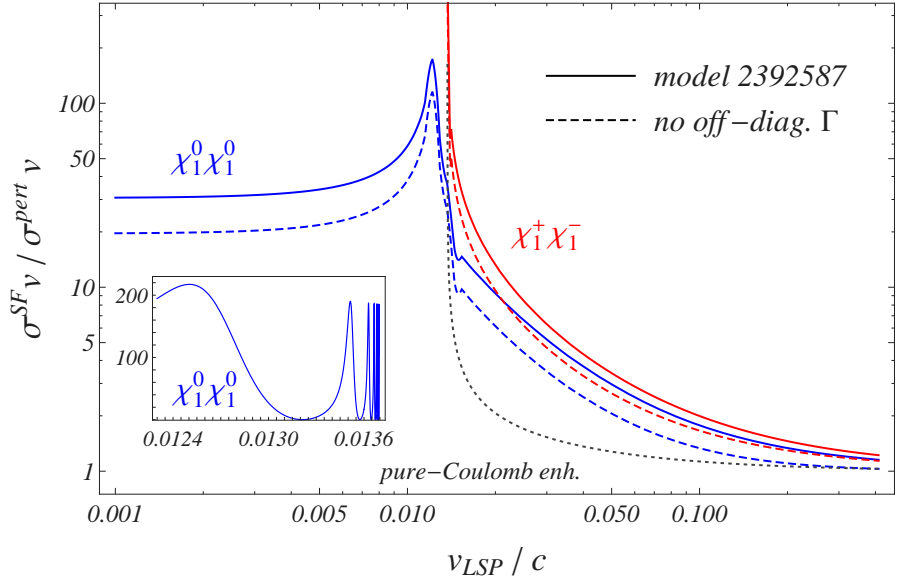}
\caption{ The enhancement of the $\chi^0_1 \chi^0_1$ and
          $\chi^+_1 \chi^-_1$ annihilation cross sections for Snowmass model
          $2392587$ relative to the perturbative tree-level rate,
          $(\sigma^\text{SF} v)/ (\sigma^\text{pert} v)$.
          The solid lines refer to the calculation of the Sommerfeld-enhanced
          rates with off-diagonal entries in the annihilation
          matrices $\Gamma$ properly included. The dashed curves show 
the enhancement with respect to the perturbative cross sections when 
off-diagonal annihilation rates are not considered. The dotted curve 
labelled ``pure--Coulomb enh.'' shows the enhancement from photon exchange 
only in the $\chi^+_1\chi^-_1$ channel. 
}
\label{fig:pMSSM_2392587_sigmavoff_Coulomb}
\end{center}
\end{figure}
%---------------------------------------------------------------------------

In Fig.~\ref{fig:pMSSM_2392587_sigmavoff_Coulomb} we plot the enhancement
$(\sigma^\text{SF} v)/(\sigma^\text{pert} v)$ of annihilation rates including
long-range  interactions, $\sigma^\text{SF} v$, with respect to the
perturbative tree-level result, $\sigma^\text{pert} v$, for the two-particle
states $\chi^0_1\chi^0_1$ and $\chi^+_1\chi^-_1$ in the
neutral sector of the model as a function of the velocity
$v_\text{LSP}$ 
of the incoming $\chi^0_1$'s in their centre-of-mass frame.
We define the velocity $v_\text{LSP}$ by
$\sqrt{s}=2m_{\chi^0_1}+m_{\chi^0_1} v_\text{LSP}^2$
with $\sqrt{s}$ the available centre-of-mass energy.
The spin-averaged tree-level annihilation rates $\sigma^\text{pert} v$ are
calculated in the non-relativistic approximation
\begin{align}
 \sigma^\text{pert} v \ = \ a  +  b  \,v^2 \, + \, \mathcal O(v^4) \ , 
\end{align}
where $v$ denotes the relative velocity of the annihilating particles.
In case of the $\chi^0_1\chi^0_1$ state the relation between the relative
velocity $v$ and $v_\text{LSP}$ is given by $v = 2\,v_\text{LSP}$.
For $\chi^+_1\chi^-_1$ annihilation reactions the relation is 
\begin{equation}
v = 2\,\mbox{Re}\,\sqrt{m_{\chi_1^0}/m_{\chi^+_1} 
[v_\text{LSP}^2 - 2\,\delta m/m_{\chi_1^0}] }\,.
\end{equation}
The coefficients $a$ and $b$ are determined from the absorptive
part of partial-wave decomposed Wilson coefficients given  in
\cite{Beneke:2012tg, Hellmann:2013jxa}.
In case of the Sommerfeld-enhanced rates $\sigma^\text{SF} v$
each partial wave contribution gets multiplied by an enhancement factor
related to the two-particle wave-function of the respective incoming state,
see \cite{paperIII} for the detailed expression. Unless otherwise stated,
Sommerfeld-enhanced results include the one-loop
corrections from heavy $\chi\chi$-states in the last potential loop, following
the approximation discussed in \cite{paperIII}.
The results for the wino-like pMSSM scenario hence include perturbative
corrections
from heavy $\chi\chi$-pairs involving the higgsino-like $\chi^0_{3,4}$ and
$\chi^\pm_2$ particles. The effects of the latter nevertheless amount
only to a negligible 
per mil level deviation on $\sigma^\text{SF}v$. This can be traced
back to the fact that the higgsino states lie at the rather high mass scale
of around $3.9\,$TeV and thus are basically decoupled.
The $(\sigma^\text{SF} v)/(\sigma^\text{pert} v)$ curves in
Fig.~\ref{fig:pMSSM_2392587_sigmavoff_Coulomb} show 
some characteristic features, which we describe next.
As there is a small mass splitting between the $\chi^0_1$ and the $\chi^\pm_1$,
the threshold for the on-shell production of the heavier neutral state
$\chi^+_1\chi^-_1$ opens at $v_\text{LSP}/c\simeq 0.014$. Well below this
threshold, the enhancement for the $\chi^0_1\chi^0_1$ system is
velocity-independent and of $\mathcal O(10)$. This saturation  
effect is characteristic for Yukawa-type interactions in the kinematic 
regime where the relative momentum of the incoming
state is well below the mass scale of the mediator: this is the case for the
$\chi^0_1\chi^0_1$ state at very small velocities, where off-diagonal Yukawa
potentials are generated by $W$-boson exchange with
$m_{\chi^0_1}\,v_\text{LSP}\ll m_W$. The
actual strength of the enhancement is, however, a combined effect of the off-diagonal 
Yukawa potential from $W$-exchange that allows for $\chi^0_1\chi^0_1\to \chi^+_1\chi^-_1$
transitions and the QED Coulomb
interaction in the (kinematically closed) $\chi^+_1\chi^-_1$
channel.
At velocities $v_\text{LSP}$ just below the $\chi^+_1\chi^-_1$ threshold
resonances in the $\chi^0_1\chi^0_1$ channel can be observed.
While the main plot in Fig.~\ref{fig:pMSSM_2392587_sigmavoff_Coulomb} displays
a curve smoothed over this region, we show in the small sub-figure a close-up
of the resonance pattern.
The existence of resonance enhancements at the threshold of a
heavier channel is well-known and has been described for instance in
\cite{Slatyer:2009vg}. However, opposed to the pattern in the close-up in
Fig.~\ref{fig:pMSSM_2392587_sigmavoff_Coulomb} no oscillating
behaviour was found in \cite{Slatyer:2009vg}, as only Yukawa potentials 
were considered.
In fact the oscillatory pattern is related to the photon exchange
in the $\chi^+_1\chi^-_1$ subsystem.
Going to even larger velocities, above the $\chi^+_1\chi^-_1$ threshold, the
enhancement in the $\chi^0_1\chi^0_1$ channel decreases, approaching one as we
depart from the non-relativistic regime.
Turning to the enhancement in the $\chi^+_1\chi^-_1$ channel, it shows quite a
different behaviour right above its threshold compared to the $\chi^0_1\chi^0_1$
system at small velocities: instead of approaching a constant value,
the enhancement factor for $\chi^+_1\chi^-_1$ rises increasingly 
as the velocities of the $\chi^\pm_1$ get smaller.
Such a behaviour is expected in the presence of long-range Coulomb-potential
interactions, where the enhancement does not saturate because the
mediator is massless. Indeed, the photon exchange between the
charged constituents of the neutral $\chi^+_1\chi^-_1$ pair dominates the
potential interactions in the regime of very small velocities:
the Yukawa potentials become very short-ranged and thus negligible compared to
the Coulomb-interaction.
The dotted (black) curve in Fig.~\ref{fig:pMSSM_2392587_sigmavoff_Coulomb}
displays the enhancement factor in the $\chi^+_1\chi^-_1$ system arising from 
Coulomb interactions due to photon exchange only. For small velocities the
pure-Coulomb enhancement factor diverges as $1/v_{\chi^+_1}$.
The true enhancement curve, that
involves all potential interactions affecting the $\chi^+_1\chi^-_1$ system
asymptotically reaches this Coulomb-like behaviour for velocities
directly above the $\chi^+_1\chi^-_1$ threshold.\footnote{Note that in spite 
of the $\propto 1/v_{\chi^+_1}$ divergence, the enhanced cross sections lead 
to a finite result in the average over the thermal velocity distribution due 
to the $v_{\chi^+_1}^2$ term in the integration measure,
$\int_{\mathbb{R}^3} d^3 \vec{v}_{\chi^+_1} = \int d\Omega\int_0^\infty dv_{\chi^+_1}\,v_{\chi^+_1}^2$.}
For larger velocities in the $\chi^+_1\chi^-_1$ system the presence of the
Yukawa potentials leads to a larger enhancement than in case of
Coulomb interactions only.

The dashed curves in Fig.~\ref{fig:pMSSM_2392587_sigmavoff_Coulomb} show
the enhancements $(\sigma^\text{SF}v)/(\sigma^\text{pert}v)$ for the 
$\chi^0_1\chi^0_1$ and $\chi^+_1\chi^-_1$ states when off-diagonal 
terms in the annihilation matrices
are (incorrectly) left out. This can lead to a $\lesssim 30\%$
underestimation of the actual enhancement in the $\chi^0_1\chi^0_1$ channel.
The effect is less pronounced for the $\chi^+_1\chi^-_1$ channel, as in this
case the cross section also gets significant contributions from ${}^3S_1$ 
annihilations and not just from ${}^1S_0$ ones.
As the ${}^3S_1$ sector is purely diagonal, the effect of
off-diagonals, relevant in the case of ${}^1S_0$ wave annihilations, becomes milder
for the spin-averaged total cross section $\sigma^\text{SF} v$.
It is worth to stress that the overall order of magnitude of the enhancements
is $\mathcal O(10)$, and becomes $\mathcal O(10^2)$ in the
resonance region around the $\chi^+_1\chi^-_1$ threshold.

%---------------------------------------------------------------------------
\begin{figure}[t]
\begin{center}
\includegraphics[width=0.75\textwidth]{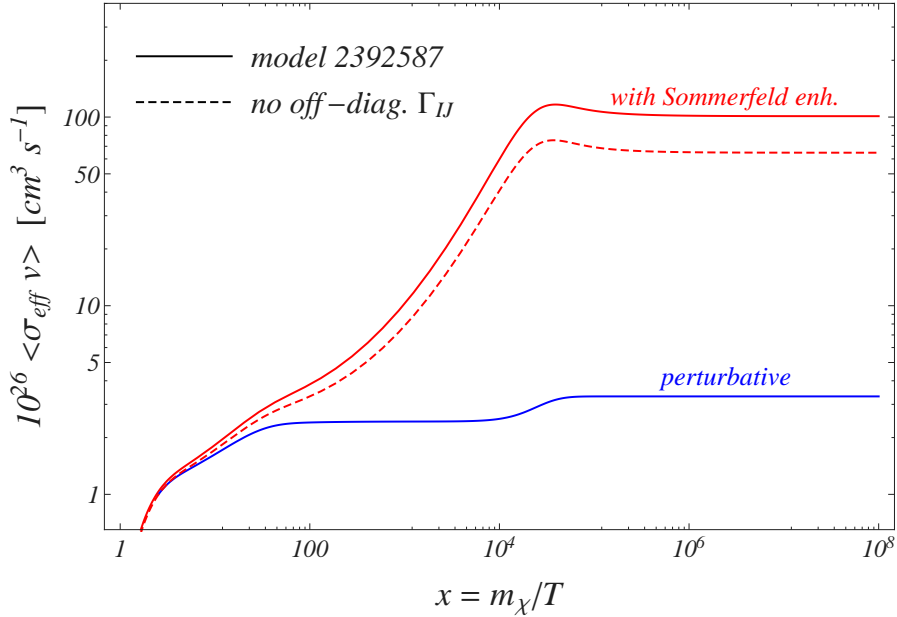}
\caption{ The thermally averaged effective annihilation rate
          $\langle \sigma_\text{eff} v \rangle$ as a function of the scaled
          inverse temperature $x = m_{\chi^0_1}/T$ in case of Snowmass model
          $2392587$. The two upper (red) curves correspond to the
          Sommerfeld-enhanced annihilation cross sections including
          (solid line) or neglecting (dashed line) the off-diagonals in the annihilation matrices.
          The lower (blue) curve represents $\langle \sigma_\text{eff} v \rangle$
          obtained from perturbative (tree-level) cross sections.
       }
\label{fig:pMSSM_2392587_sigmaeffoff}
\end{center}
\end{figure}
%---------------------------------------------------------------------------

The quantity that enters the Boltzmann equation for the neutralino 
number density is the thermally averaged
effective annihilation rate $\langle \sigma_\text{eff} v\rangle$. 
Fig.~\ref{fig:pMSSM_2392587_sigmaeffoff} shows 
$\langle \sigma_\text{eff} v\rangle$ as defined in \cite{paperIII}
as a function of the inverse scaled temperature $x = m_{\chi^0_1}/T$.
The lower solid (blue) curve represents the perturbative (tree-level)
annihilation rates while the upper solid and the dashed (red) lines refer to
Sommerfeld-enhanced cross sections including and neglecting off-diagonal 
annihilation rates, respectively. The plot can be
divided into several regions with different characteristics. Let us first note
that for $x \lesssim 10$ the depicted behaviour of
$\langle \sigma_\text{eff} v\rangle$ is unphysical.
The mean velocity of the annihilating particles
in the plasma scales as $\sqrt{1/x}$ and hence is no longer non-relativistic for
$x \lessim 10$ while the results of our framework strictly apply only to 
non-relativistic $\chi\chi$ pair-annihilations, {\it i.e.} for $x\gtrsim10$.
Around $x\sim20$ the annihilation rates of $\chi^0_1$ and $\chi^+_1$ can no
longer maintain chemical equilibrium and the
particles start to decouple from the thermal plasma. Hence only the region above
$x\sim20$ is important for the
calculation of the relic abundance. Around $x\gtrsim10^4$ the number densities
of the $\chi^\pm_1$ are so strongly Boltzmann suppressed with respect to the
$\chi^0_1$ number density despite the small mass splitting 
that the rates of the charginos basically play no
role in the effective rate $\langle \sigma_\text{eff} v \rangle$, which is then
essentially given by $\chi^0_1\chi^0_1$ annihilations. Note
that we can estimate the point of chargino decoupling between $x\sim10^4-10^5$
from the ratio of the Boltzmann distributions
$n_{\chi^+_1}/n_{\chi^0_1}\propto \exp(-\delta m/m_{\chi^0_1}\,x)$, taking the
$\mathcal O(10^{-1}\,$GeV$)$ mass splitting into account.
After $\chi^\pm_1$ decoupling, $\langle \sigma_\text{eff} v\rangle$ including the
Sommerfeld enhancements becomes constant, which we can infer from the constant
enhancement factor for the $\chi^0_1\chi^0_1$ system for very low velocities
shown in Fig.~\ref{fig:pMSSM_2392587_sigmavoff_Coulomb}.
Before $\chi^\pm_1$ decoupling, $\langle \sigma_\text{eff} v\rangle$ including
the Sommerfeld enhancements rises with increasing $x$ due to the contributions
from the charginos but also due to the velocity-dependent enhancement on the
$\chi^0_1\chi^0_1$ system itself for larger relative velocities.
On the contrary, the perturbatively determined
$\langle \sigma_\text{eff} v \rangle$ shows a constant behaviour before
and after $\chi^\pm_1$ decoupling with a rise only around the decoupling region;
the contributions that dominate the perturbative cross sections in the
non-relativistic regime are the velocity-independent leading-order $S$-wave
terms.

%---------------------------------------------------------------------------
\begin{figure}[t]
\begin{center}
\includegraphics[width=0.75\textwidth]{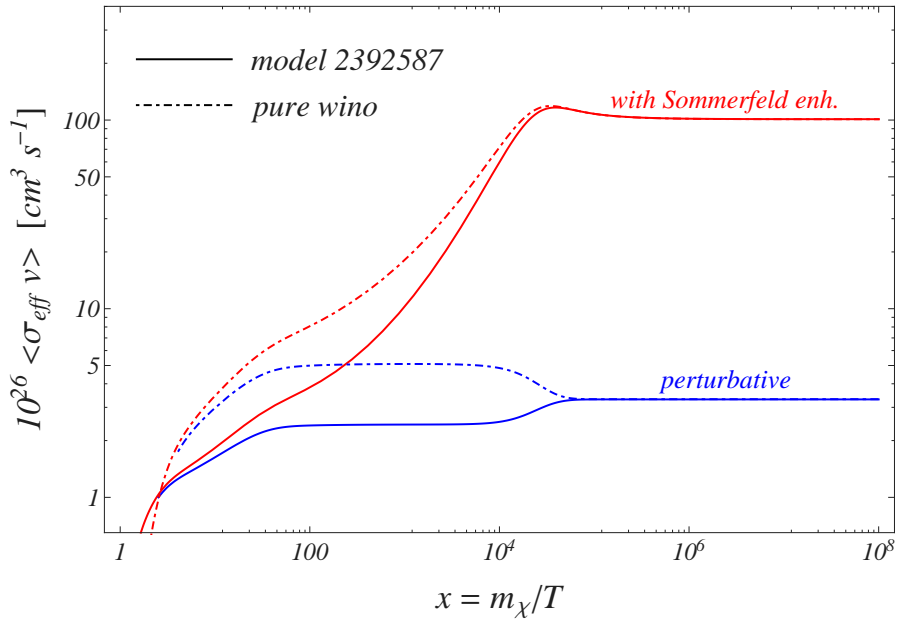}
\caption{ Thermally averaged effective annihilation rates
         $\langle \sigma_\text{eff} v \rangle$ as a function of the scaled
         inverse temperature $x = m_{\chi^0_1}/T$. The two upper (red) curves
         refer to a calculation with Sommerfeld-enhanced cross sections while
         the two lower (blue) curves represent the perturbative results.
         Solid lines correspond to the Snowmass pMSSM scenario $2392587$ and
         dot-dashed curves show the results for the pure-wino scenario.
       }
\label{fig:pMSSM_2392587_comparepure}
\end{center}
\end{figure}
%---------------------------------------------------------------------------
Fig.~\ref{fig:pMSSM_2392587_comparepure} compares the
thermally averaged effective rates
$\langle \sigma_\text{eff} v\rangle$ as calculated from the wino-like pMSSM
scenario
and from a pure-wino $SU(2)_L$ triplet minimal dark matter model with the same
$\chi^0_1$ mass.
In the pure-wino model the mass splitting between the $\chi^0_1$ and
$\chi^\pm_1$ has to be kept in the Schr\"odinger equation as it is of the same
order as the non-relativistic kinetic energy and the potentials. However
in the hard annihilation rates the mass splitting is a subleading effect
and is neglected; the annihilation matrices in the pure-wino model depend on the
$\chi^0_1$ mass only (the corresponding expressions can be found, for instance,
in \cite{paperIII}).
While the rates for $\chi^0_1\chi^0_1$
annihilations agree at permille level, the cross sections involving $\chi^\pm_1$
are generically larger by factors of $\mathcal O(1)$ in the pure-wino model as
compared to the pMSSM wino-like model. This can be mainly traced back to the
destructive interference between $t$-channel sfermion and $s$-channel
$Z$ (and Higgs-boson) exchange amplitudes in $\chi^+_1\chi^-_1 \rightarrow ff$
annihilations in the pMSSM scenario case, while the $t$-channel sfermion
exchange amplitudes are absent in the pure-wino model.
In addition the pure-wino case neglects
all final state masses which in particular gives rise to larger annihilation
rates into the $t\overline t$ and electroweak gauge boson final states as
compared to the pMSSM scenario, where the non-vanishing masses of 
all SM particles are taken into account.
This accounts for the deviation between the curves in
Fig.~\ref{fig:pMSSM_2392587_comparepure} before $\chi^\pm_1$ decoupling.

%---------------------------------------------------------------------------
\begin{figure}[t]
\begin{center}
\includegraphics[width=0.75\textwidth]{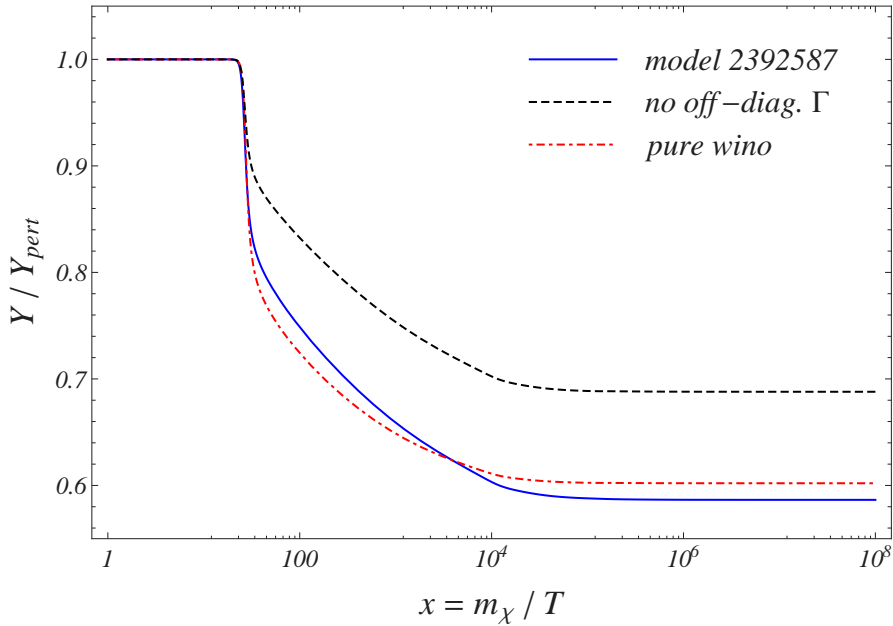}
\caption{ The ratios of the yield $Y/Y_\text{pert}$ as a function of
          $x = m_{\chi^0_1} / T$, where $Y$ is calculated including the
          Sommerfeld enhancement on the $\chi\chi$ annihilation rates while
          $Y_\text{pert}$ just uses the perturbative ones.
          The solid (blue) and dashed (black) curves give the results for the
          Snowmass model $2392587$ including and neglecting off-diagonal
          annihilation rates, respectively. The dot-dashed (red) curve
         corresponds to $Y/Y_\text{pert}(x)$ in the pure-wino model.
       }
\label{fig:pMSSM_2392587_yield}
\end{center}
\end{figure}
%---------------------------------------------------------------------------
Finally we consider the yield $Y=n/s$, defined as the ratio of the number
density $n$ of all co-annihilating particle species divided by the entropy
density $s$ in the cosmic co-moving frame.
The dependence of the yield on the scaled inverse temperature $x=m_{\chi^0_1}/T$
is governed by a Boltzmann equation and the $\chi^0_1$ relic abundance
is obtained from the yield today.
In Fig.~\ref{fig:pMSSM_2392587_yield} we show the ratio of the yield $Y$
calculated from Sommerfeld-enhanced cross sections in both the pMSSM and the
pure-wino model to the corresponding results using perturbative cross sections, 
$Y_\text{pert}$, as a function of $x$.

First note, that the denominator $Y_\text{pert}$ in the ratio
$Y/Y_\text{pert}$ differs for the pMSSM and the pure-wino model, which is a
consequence of the different effective rates
$\langle \sigma_\text{eff} v \rangle$, see
Fig.~\ref{fig:pMSSM_2392587_comparepure}.
Further, in case of the pMSSM scenario we show results corresponding to a
calculation of $Y$ including and neglecting off-diagonal annihilation rates.
Around $x\sim20$ the yields including Sommerfeld enhancements start to
depart from the corresponding perturbative results;
the enhanced rates delay the freeze-out of interactions, which
leads to a reduction of the yield $Y$ compared to the perturbative result
$Y_\text{pert}$.
The most drastic reduction in $Y/Y_\text{pert}$ occurs between $x\sim 20$ and
$x\sim 10^3$. In this region the
enhancement factors on the cross sections are of $\mathcal O(10)$ (and not yet
$\mathcal O(10^2)$ as for very large $x$), leading to 
$Y/Y_\text{pert}$ values that deviate  from $1$ by a few $10\%$.
For $x \gtrsim10^5$ the fraction $Y/Y_\text{pert}$ stays constant, meaning that
at these temperatures
the particle abundances in both the perturbative and Sommerfeld-enhanced
calculation are frozen in.
In case of the wino-like model we find that the relic densities calculated from 
the yield today read $\Omega^\text{pert} h^2 = 0.112$ and
$\Omega^\text{SF} h^2 = 0.066$.
Hence taking into account the Sommerfeld effect leads to a reduction of the
calculated relic abundance of around $40\%$.
On the other hand, neglecting the off-diagonal
annihilations in the calculation of Sommerfeld-enhanced rates
overestimates the relic density by $15\%$ compared to the correct 
$\Omega^\text{SF} h^2$.
Let us recall that the relic density calculated without
corrections from heavy $\chi\chi$-states in the last potential loop
differs from the $\Omega^\text{SF}h^2$ value quoted above at most at the
per mil level.
Due to overall larger hard annihilation rates in the pure-wino model, the
calculated relic density including Sommerfeld-enhanced rates turns out to be
$\Omega_{\text{pure-w}}^\text{SF} h^2 = 0.034$, while the corresponding
perturbative result is $\Omega_{\text{pure-w} }^\text{pert} h^2 = 0.056$.

It is difficult to quantify the theoretical error on such numbers. 
Conventional tree-level calculations of annihilation cross sections 
and the ensuing relic densities neglect radiative corrections and 
are supposed to be accurate to ${\cal O}(5\%)$ in the absence of 
enhanced corrections due to non-relativistic scattering, large Sudakov 
logarithms, or, potential strong-interaction effects for quark and 
gluon final states. When the Sommerfeld effect is included, the latter 
two restrictions still apply. The computation of the Sommerfeld 
effect itself neglects ${\cal O}(v^2)$ corrections to the 
scattering potentials 
as well as ordinary, non-enhanced corrections to the short-distance 
annihilation coefficients. Hence the accuracy of the Sommerfeld-corrected 
annihilation cross sections and relic densities is presumably again 
at the  ${\cal O}(5\%)$ accuracy level at best.

%-------------------------------------------------------------------------------

\section{Higgsino-like $\chi^0_1$}
\label{sec:res_higgsino}
The higgsino-like neutralino $\chi^0_1$ arises as the lightest out of four mass
eigenstates
$\chi^0_{1,2}, \chi^\pm_1$ related to two $SU(2)_L$ fermion
doublets. Note that the hypercharges of the two $SU(2)_L$ doublets are given
by $Y=\pm1/2$ respectively, which ensures the electric neutrality of
the $\chi^0_1$.
The common mass scale of the $\chi^0_{1,2},\chi^\pm_1$ states is set by the
$\mathcal O($TeV$)$ higgsino mass parameter, $m_\chi \sim \vert\mu\vert$.
Electroweak symmetry breaking introduces a tree-level splitting between
$m_{\chi^0_1}$ and the masses of the three heavier states of 
$\mathcal O(m_Z^2/m_\text{LSP})\sim\mathcal O(1\,$GeV$)$.
This is considerably larger than the tree-level mass splitting
in the wino-like $\chi^0_1$ case; in particular loop corrections play a
sub-dominant role in the mass splittings of higgsino-like neutralinos and
charginos.

As an example of this class of models we consider the
Snowmass pMSSM scenario with ID $1627006$ \cite{Cahill-Rowley:2013gca}, that
features a higgsino-like $\chi^0_1$ LSP with $m_{\chi^0_1} = 1172.31\,$GeV and
higgsino fraction $\vert Z_{31}\vert^2+\vert Z_{41}\vert^2 = 0.98$. The
heavier higgsino-like states $\chi^\pm_1$ and $\chi^0_2$ have a mass splitting of
$\delta m_{\chi^+_1} = 1.8\,$GeV and $\delta m_{\chi^0_2} = 9.5\,$GeV to the
$\chi^0_1$ mass.
Again, all pMSSM spectrum parameters are taken without any modification from the
corresponding Snowmass (slha) model-file $1627006$ provided by
\cite{Cahill-Rowley:2013gca}.

As in Sec.~\ref{sec:res_wino}, it is instructive to compare the pMSSM scenario
with higgsino-like $\chi^0_1$ and co-annihilating $\chi^0_2$ and $\chi^\pm_1$
to a model with pure-higgsino $\chi^0_{1,2}, \chi^\pm_1$ states and completely
decoupled sfermions and heavy Higgses. We refer to the latter scenario as
``pure-higgsino'' model; such model is also discussed in the 
context of Minimal Dark Matter \cite{Cirelli:2007xd}. Pure-Higgsino states 
interact only with the SM gauge bosons $W^\pm, Z, \gamma$ but not with the 
Higgs bosons. The accessible final states in $2\rightarrow2$
co-annihilation reactions of pure higgsinos are hence given by 
particle pairs formed out of SM gauge bosons and fermions as well as of the (SM-like) Higgs
$h^0$, where all these SM particles are taken to be massless,
and only SM gauge bosons and higgsinos appear as intermediate states in 
tree-level annihilations.
The co-annihilation rates of the higgsino-like $\chi^0_{1,2}, \chi^\pm_1$ states
in the pMSSM scenario $1627006$ happen to be larger than the corresponding
reactions in the pure-higgsino case. This can be traced back to the presence of
non-decoupled sfermion and Higgs states in the higgsino-like $\chi^0_1$ pMSSM
model and in particular to non-decoupled wino-like states
$\chi^0_3, \chi^\pm_2$ at the scale of $1.6$~TeV.

In the determination of the $\chi^0_1$ relic abundance for this pMSSM scenario
including co-annihilations only the higgsino-like states are
relevant. Other heavier states are already sufficiently
Boltzmann-suppressed during $\chi^0_1$ freeze-out. Hence we neglect
the co-annihilations of the lightest sfermion states
$\tilde \tau_1$ and $\tilde\nu_3$, with masses around $1.44\,$TeV,
although we include co-annihilation reactions of all heavier $\chi^0/\chi^\pm$ states.
Yet the latter have basically no effect on the $\chi^0_1$ relic density, as
their abundances are already sufficiently suppressed at $\chi^0_1$ decoupling.
Obviously, in the pure-higgsino scenario only the co-annihilations between  
the higgsino-like
species  $\chi^0_{1,2},\chi^\pm_1$
are taken into account for the calculation of the  relic abundance.

We consider Sommerfeld corrections to all co-annihilation rates between two
higgs\-ino-like particles in both the pMSSM scenario $1627006$ and the
pure-higgsino model by treating all channels
built from the states $\chi^0_{1,2}, \chi^\pm_1$ exactly in the corresponding
Schr\"odinger equations.
Moreover, the remaining heavier $\chi^0/\chi^\pm$ two-particle states in the
higgsino-like pMSSM scenario are treated perturbatively in the last potential
loop, see \cite{paperIII}.
In case of the pure-higgsino model though, all heavier states are considered
as completely decoupled.
Dividing the co-annihilation reactions 
into sets corresponding to total electric charge,
we identify a neutral sector with the four two-particle states
$\chi^0_1\chi^0_1, \chi^0_1\chi^0_2, \chi^0_2\chi^0_2$ and $\chi^+_1\chi^-_1$. The
single-positive (negative) charged sector contains the two states
$\chi^0_1\chi^+_1, \chi^0_2\chi^+_1$ ($\chi^0_1\chi^-_1, \chi^0_2\chi^-_1$),
whereas the double-positive (double-negative)
charged sector features only one two-particle state relevant in co-annihilations
with the higgsino-like $\chi^0_1$ dark matter candidate: $\chi^+_1\chi^+_1$
($\chi^-_1\chi^-_1$). Note that annihilations of the latter double-charged states
$\chi^+_1\chi^+_1$ and $\chi^-_1\chi^-_1$ are absent in the pure-higgsino model
due to hypercharge conservation in this $SU(2)_L\times U(1)_Y$ symmetric limit,
as they have a non-zero hypercharge, namely $Y_{\chi^\pm\chi^\pm} = \pm 1$. 
In contrast, in the higgsino-like
$\chi^0_1$ pMSSM case with broken $U(1)_Y$ symmetry, annihilations of the
double-charged channels into a $W^+W^+$ or $W^-W^-$ pair are possible, though the
rates are suppressed by a factor $\sim m_W/m_{\chi^0_1}$ compared to the
magnitude of the neutral sector's leading rates.

%---------------------------------------------------------------------------
\begin{figure}[t]
\begin{center}
\includegraphics[width=0.75\textwidth]{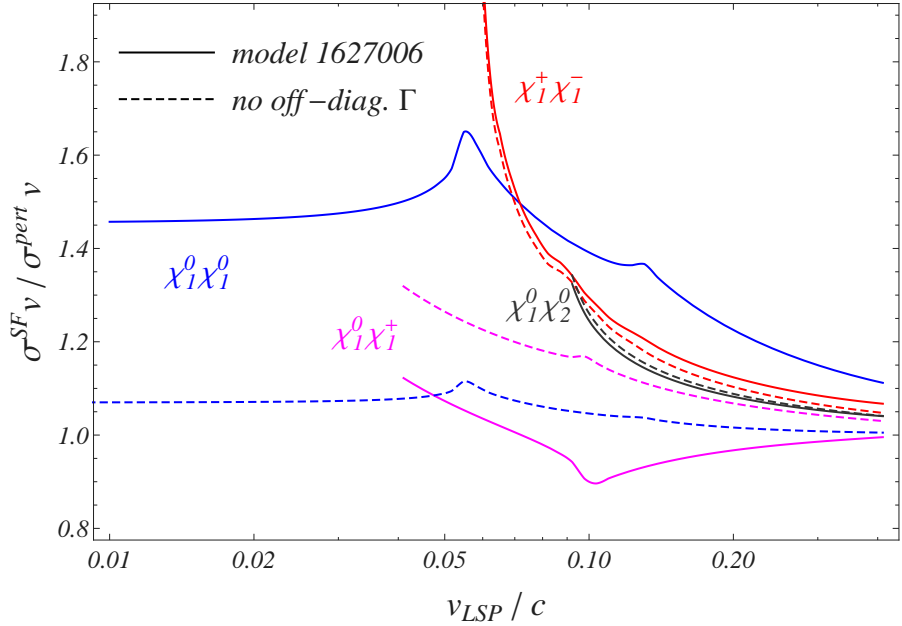}
\caption{ Enhancement factors $(\sigma^\text{SF} v)/ (\sigma^\text{pert} v)$
          in the four most relevant two-particle channels $\chi^0_1 \chi^0_1$,
          $\chi^+_1\chi^-_1$, $\chi^0_1 \chi^0_2$ and $\chi^0_1 \chi^+_1$
          of Snowmass model $1627006$.
          The enhancement factor for the additionally relevant channel
          $\chi^0_1\chi^-_1$ agrees with the one for the $\chi^0_1\chi^+_1$ pair.
          Solid lines refer to the calculation of the Sommerfeld-enhanced
          rates with off-diagonal terms in the annihilation
          matrices properly included. Dashed curves show the enhancement when 
          the off-diagonal annihilation rates are  neglected.
       }
\label{fig:pMSSM_1627006_sigmavoff_Coulomb}
\end{center}
\end{figure}
%---------------------------------------------------------------------------
Fig.~\ref{fig:pMSSM_1627006_sigmavoff_Coulomb} shows the enhancement
$(\sigma^\text{SF} v)/(\sigma^\text{pert} v)$ of the individual cross sections
for those channels that have the most relevant contribution to the 
relic abundance
calculation, that is $\chi^0_1\chi^0_1,\, \chi^+_1\chi^-_1,\,
\chi^0_1\chi^0_2$ in the neutral sector, and $\chi^0_1\chi^+_1$ in the
single-charged sector ($\chi^0_1\chi^-_1$ gives the same contribution). 
First note that the enhancements are only of $\mathcal O(1)$,
opposed to $\mathcal O(10^2)$ enhancements in case of the wino-like model in
Sec.~\ref{sec:res_wino}. This can be explained due to the larger mass splittings
to the next-to-lightest states $\chi^\pm_1$, $\chi^0_2$ in the higgsino-like
$\chi^0_1$ case and the fact that
the couplings to SM gauge bosons and (light) Higgs particles are generically
smaller for higgsinos than for winos.
The enhancement of the $\chi^0_1\chi^0_1$ rate as a function of the velocity
$v_\text{LSP}$ shows again the saturated, velocity-independent behaviour typical
for Yukawa type potentials in the low velocity regime well below the thresholds
of the heavier two-particle states.
As in the wino-model, both the off-diagonal Yukawa potential and the (diagonal)
Coulomb potential in the kinematically closed $\chi^+_1\chi^-_1$ channel
contribute here to the actual size of the enhancement.
At larger velocities, two
resonance regions at the thresholds for $\chi^+_1\chi^-_1$ and $\chi^0_2\chi^0_2$
production are visible 
(the $\chi^0_2\chi^0_2$ channel opens up at $v_{\rm LSP}/c\simeq 0.127$;
the ratio $(\sigma^\text{SF} v)/(\sigma^\text{pert} v)$ for this channel is
very close to 1, and is not shown in Fig.~\ref{fig:pMSSM_1627006_sigmavoff_Coulomb}).
One might ask why no resonance at the $\chi^0_1\chi^0_2$ threshold is visible in the
$\chi^0_1\chi^0_1$ channel:
recall that Fermi-statistics forbids the $\chi^0_1\chi^0_1$-pair to build the
totally symmetric partial-wave configurations ${}^3S_1$ and ${}^1P_1$.
In case of unbroken $SU(2)_L\times U(1)_Y$ symmetry it turns out, though,
that the $\chi^0_1\chi^0_2$ pair can build ${}^3S_1$ and ${}^1P_1$ configurations
but not ${}^1S_0$ and ${}^3P_{\mathcal J}$ states. Hence there are no
off-diagonal entries in the neutral potential matrices 
encoding $\chi^0_1\chi^0_1 \leftrightharpoons \chi^0_1\chi^0_2$ interactions
in the pure-higgsino limit.
Departing from the $SU(2)\times U(1)_Y$ symmetric limit
gives rise to $\chi^0_1\chi^0_2$ contributions to the enhancement
$(\sigma^\text{SF} v)/(\sigma^\text{pert} v)$ in the $\chi^0_1\chi^0_1$ channel that
are however suppressed by $(m_W/m_{\chi^0_1})^3$ with respect to the leading
contributions; this explains why no $\chi^0_1\chi^0_2$ threshold effect is visible in
Fig.~\ref{fig:pMSSM_1627006_sigmavoff_Coulomb}.
Such restrictions due to non-accessible partial-wave configurations do not
exist for the next-to-lightest neutral two-particle state
$\chi^+_1\chi^-_1$, and resonances at the thresholds of all co-annihilating
neutral $\chi\chi$-pairs heavier than the $\chi^+_1\chi^-_1$ are
visible in the latter channel in
Fig.~\ref{fig:pMSSM_1627006_sigmavoff_Coulomb}. Furthermore, note
the $1/v_{\chi^+_1}$ Coulomb-type enhancement in the
$\chi^+_1\chi^-_1$ channel directly above its threshold caused by potential
photon-exchange between the $\chi^+_1$ and $\chi^-_1$. The Coulomb potential
surpasses the potentials from massive gauge boson and Higgs exchange at very
small velocities in the $\chi^+_1\chi^-_1$ channel, but for moderate velocities
both the Coulomb and the (off-)diagonal Yukawa interactions are relevant.
Turning to channel $\chi^0_1\chi^0_2$, the
corresponding enhancement $(\sigma^\text{SF} v) / (\sigma^\text{pert} v)$ 
increases as the velocity decreases. In particular,
there is no saturation of the enhancement directly above threshold,
because the lighter
channels $\chi^0_1\chi^0_1$ and especially $\chi^+_1\chi^-_1$ are always
kinematically open and accessible from an on-shell $\chi^0_1\chi^0_2$ 
state via off-diagonal potential interactions. 

The ratio
$(\sigma^\text{SF} v)/(\sigma^\text{pert} v)$ for the charged state $\chi_1^0\chi_1^+$
that is additionally plotted in Fig.~\ref{fig:pMSSM_1627006_sigmavoff_Coulomb} 
(lowermost magenta line) shows
that the Sommerfeld effect can also produce corrections that
reduce the perturbative result. For the channel $\chi_1^0\chi_1^+$ the negative
correction arises from the interference of amplitudes where,
after multiple electroweak and Higgs boson exchanges, the state that 
annihilates into the light final state particles is the same as the incoming one, $\chi_1^0\chi_1^+$,
with amplitudes where
the actual state that annihilates is $\chi_2^0\chi_1^+$. In the EFT formalism such interferences
arise from the off-diagonal annihilation terms
$\chi_1^0\chi_1^+ \to \chi_2^0 \chi_1^+$
and $\chi_2^0\chi_1^+ \to \chi_1^0 \chi_1^+$, combined with the
off-diagonal potential term for $\chi_1^0\chi_1^+ \to \chi_2^0 \chi_1^+$.
The dashed magenta curve in Fig.~\ref{fig:pMSSM_1627006_sigmavoff_Coulomb}
refers to the situation where off-diagonal short-distance rates are neglected
in the calculation of the Sommerfeld enhanced $\chi^0_1\chi^+_1$ annihilation
cross section. It is nicely seen that the destructive interference effect
disappears in this case and the ratio
$(\sigma^\text{SF} v)/(\sigma^\text{pert} v)$  is always positive.
The enhancement in the $\chi_1^0\chi_1^+$ channel also saturates as its
on-shell production threshold is approached.
This should be the case as the $\chi^0_1\chi^+_1$ channel is the lightest
in the single positive-charged sector, and its behaviour should be similar to
the one of the lightest neutral channel, $\chi^0_1\chi^0_1$, directly above
threshold. However, such saturation is not visible in
Fig.~\ref{fig:pMSSM_1627006_sigmavoff_Coulomb}
because there we plot the $\chi_1^0\chi_1^+$  cross section as a function 
of $v_{\rm LSP}$ and not as a function of the relative velocity of the 
channel, related to the latter by 
$v^2= 2(m_{\chi_1^0}+m_{\chi_1^+})/(m_{\chi_1^0}m_{\chi_1^+})\times 
(m_{\chi_1^0}v_{\rm LSP}^ 2-2\delta m)$.\footnote{If the 
$\chi_1^0\chi_1^+$ cross section behaves
as $\sigma^\text{SF} v \simeq a +b v^2$ close to threshold, the saturation is visible because of the zero slope of this
function at $v=0$; in terms of $v_{\rm LSP}$ it 
reads $\sigma^\text{SF} v = a +b^\prime (v_{\rm LSP}^2 - c)$, which does not have a zero slope
at the threshold of the channel, $v_{\rm LSP}=\sqrt{c}$.}
Let us also mention that the
dip  in the $\chi_1^0\chi_1^+$ cross section caused by interference effects
is located at the velocity where the other state included in the Sch\"odinger
equation for this charge sector, $\chi_2^0 \chi_1^+$, opens up.

As we have already noted in context of the $\chi^0_1\chi^+_1$ channel above,
the dashed curves in Fig.~\ref{fig:pMSSM_1627006_sigmavoff_Coulomb}
show the results for the corresponding enhancements 
of the pMSSM scenario $1627006$ when off-diagonal
annihilation rates are neglected.
This disregard would lead to an underestimation of the actual enhancement due
to the long-range potential
interactions of around $30\%$ in the $\chi^0_1\chi^0_1$ channel.
%and around $10\%$  in the $\chi^0_1\chi^+_1$ channel.
The effect is much milder for the $\chi^0_1\chi^0_2$ and $\chi^+_1\chi^-_1$
pairs and is explained by the contributions of ${}^3S_1$ partial-wave
annihilations to the cross sections (absent for the identical particle-pair
channel $\chi^0_1\chi^0_1$); off-diagonal ${}^3S_1$ annihilation
rates are suppressed relative to the leading (diagonal) rates by an order of
magnitude, due to destructive interference effects between sfermion and
gauge boson exchange amplitudes. Hence, as off-diagonals play a minor role
in ${}^3S_1$ annihilations, their effect in the spin-averaged cross sections
$\sigma^\text{SF} v$ will also be less pronounced.
As the conclusions on the enhancements in case of the pure-higgsino
$\chi^0_1$ model are similar to the results
in Fig.~\ref{fig:pMSSM_1627006_sigmavoff_Coulomb} we do not show a corresponding
plot here. Let us mention though again, that the hard co-annihilation
rates in the pure-higgsino model are a few percent smaller than in the
higgsino-like $\chi^0_1$ model. Furthermore, the off-diagonal rates for
${}^3S_1$ annihilations in the system of $\chi^0_1\chi^0_2$ and $\chi^+_1\chi^-_1$
states are of the same order of magnitude as the diagonal ones.

%---------------------------------------------------------------------------
\begin{figure}[t]
\begin{center}
\includegraphics[width=0.74\textwidth]{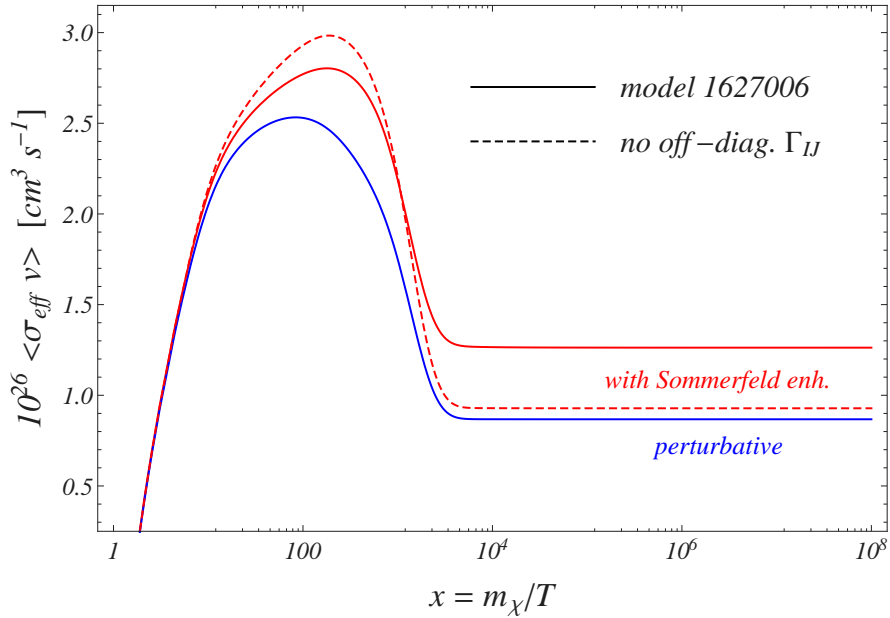}
\caption{ The thermally averaged effective annihilation rate
          $\langle \sigma_\text{eff} v \rangle$ as a function of the scaled
          inverse temperature $x = m_{\chi^0_1}/T$ for the pMSSM Snowmass
          model $1627006$ with higgsino-like $\chi^0_1$.
          The upper two (red) curves refer to the 
          Sommerfeld-enhanced cross sections: the solid line 
          includes the off-diagonal annihilation rates while the dashed curve  
          does not.
          The lowermost (blue) curve corresponds to the perturbative result.
       }
\label{fig:pMSSM_1627006_sigmaeff}
\end{center}
\end{figure}
%---------------------------------------------------------------------------
Fig.~\ref{fig:pMSSM_1627006_sigmaeff} shows the
thermally averaged effective annihilation rate
$\langle \sigma_\text{eff} v\rangle$ as a function of the inverse scaled
temperature $x$. The lower solid (blue) curve
represents the result using perturbatively calculated rates, while the upper two
(red) curves with solid and dashed line style refer to computations with
Sommerfeld-enhanced cross sections including and neglecting off-diagonal
annihilation rates, respectively. Again the
region for $x \lesssim 10$ is unphysical, as the co-annihilating particles'
mean velocities are
outside the non-relativistic regime. Due to larger mass splittings between the
higgsino-like neutralino and chargino states, the decoupling of the heavier
states $\chi^\pm_1$ and $\chi^0_2$ takes place already around $x\simeq10^3$.
As can be seen from Fig.~\ref{fig:pMSSM_1627006_sigmaeff},
the Sommerfeld effect enhances the thermally averaged effective annihilation
cross section by $3\%$ up to $25\%$ with respect to the perturbative result
in the region of $x$ around $10-10^3$ which is most relevant in the relic
abundance calculation.
The effect of correctly treating off-diagonal annihilation rates
is most essential for large values of $x$ in the range $10^4 - 10^8$, where
$\langle\sigma_\text{eff}v\rangle$ would be underestimated by around $25\%$ if
off-diagonals were neglected in the hard annihilation rates. 
In the region $x=10-10^3$ the effect of off-diagonal rates is
also noticeable, leading to an overestimation of
$\langle\sigma_\text{eff}v\rangle$ that reaches $6\%$ if off-diagonal
rates are not taken into account. The latter difference with respect to the
true result is traced back to the contribution to 
$\langle \sigma_\text{eff} v\rangle$ of the 
charged $\chi_1^0 \chi_1^+$ channel, which in the absence
of off-diagonal annihilation terms does not 
get the negative interference term that lowers the Sommerfeld-corrected
cross section, see Fig.~\ref{fig:pMSSM_1627006_sigmavoff_Coulomb}.
Once the $\chi^\pm_1$ particles are decoupled, the contributions of the 
channels $\chi_1^0 \chi_1^\pm$ to $\langle \sigma_\text{eff} v\rangle$ basically
vanish. The much larger enhancement in the $\chi_1^0 \chi_1^0$
cross section when off-diagonal rates are consistently taken into account then
explains why the correct $\langle \sigma_\text{eff} v\rangle$ result crosses
the dashed line for $x\grtsim 10^3$ in Fig.~\ref{fig:pMSSM_1627006_sigmaeff}.

%---------------------------------------------------------------------------
\begin{figure}[t]
\begin{center}
\includegraphics[width=0.75\textwidth]{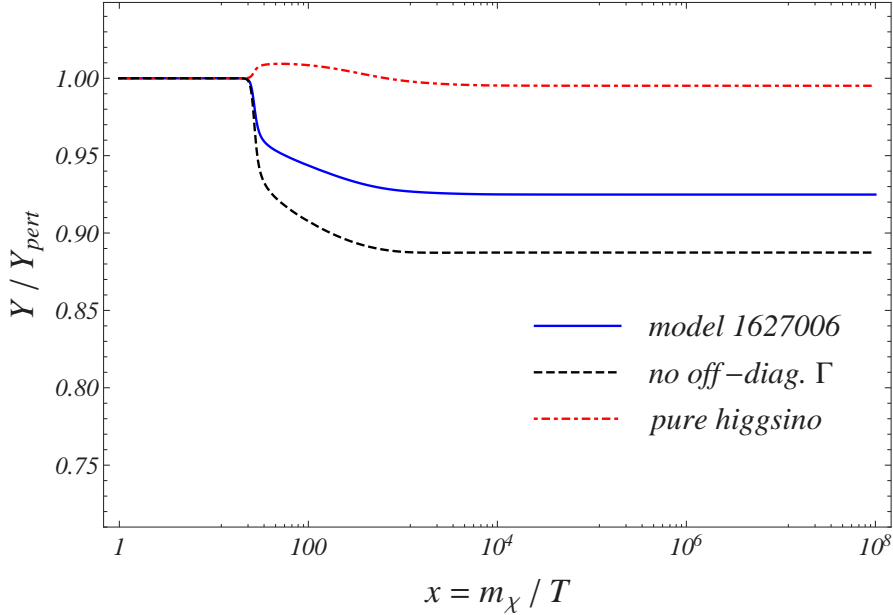}
\caption{The ratio of the yields $Y/Y_\text{pert}$, where
         $Y$ is calculated including Sommerfeld enhancements on the
         annihilation rates and $Y_\text{pert}$ uses 
         purely perturbative rates. The solid (blue) line corresponds to the
         calculation within the pMSSM Snowmass model $1627006$ with
         higgsino-like $\chi^0_1$, that includes off-diagonal
         annihilation rates. The dashed (black) line gives the same result but
         with neglected off-diagonal rates. The dot-dashed curve
         is the result (with off-diagonal rates) obtained for the 
         pure-higgsino model.
       }
\label{fig:pMSSM_1627006_Y}
\end{center}
\end{figure}
%---------------------------------------------------------------------------

Finally, Fig.~\ref{fig:pMSSM_1627006_Y} shows the ratio $Y/Y_\text{pert}$.
The solid (blue) and dashed (black) curves refer to calculations within the
pMSSM Snowmass model $1627006$ with off-diagonal annihilation reactions
included and neglected, respectively.
The dot-dashed (red) line applies to the pure-higgsino model. 
The relic abundances that we calculate within the pMSSM Snowmass model read
$\Omega^\text{pert} h^2 = 0.108$ if perturbative annihilation reactions are
considered and $\Omega^\text{SF} h^2 = 0.100$ taking Sommerfeld-enhanced rates
into account. Accounting for the long-range potential
interactions hence leads to a reduction of $8\%$ on the predicted relic
density for the pMSSM higgsino-like $\chi^0_1$ model.
Neglecting off-diagonal rates in the pMSSM Snowmass model calculation
reduces the relic abundance to a value $\Omega^\text{SF,\,no-off} h^2 = 0.096$.
This is because the
effective thermal average cross section without the off-diagonal rates
is larger in the region where
chemical decoupling takes place, see Fig.~\ref{fig:pMSSM_1627006_sigmaeff}.
The error on
$\Omega^\text{SF} h^2$ when disregarding off-diagonal rates therefore amounts to
an underestimation of $4\%$ in this case.
The Sommerfeld-enhanced rates without the one-loop corrections from
heavy $\chi\chi$-states in the last potential loop before annihilation give
a $1\%$ deviation on the final $\Omega^\text{SF} h^2$ result.
In contrast, the relic abundances in the pure-higgsino model,
obtained using perturbative or Sommerfeld-enhanced rates, almost
coincide, namely $\Omega^\text{pert}_{\text{pure-h}}h^2 = 0.127$ and
$\Omega^\text{SF}_{\text{pure-h}}h^2 = 0.126$, where the latter result 
includes the off-diagonal rates. As can be expected, the overall smaller
annihilation rates in the pure-higgsino scenario lead to a larger relic
abundance than in the higgsino-like pMSSM scenario.
The fact that the perturbative yield surpasses the Sommerfeld-corrected one
right after chemical decoupling in the
pure-higgsino model is explained by the slightly smaller
$\langle\sigma_\text{eff}v\rangle$ in the Sommerfeld-corrected result
in that region of $x$, which is in turn produced by the 
Sommerfeld suppression in the charged channels $\chi^0_1 \chi^\pm_1$. 
Overall, there is a strong cancellation between cross section enhancement in 
the neutral and suppression in the charged channels, leading to an 
almost vanishing net Sommerfeld correction.

\section{Light scenario}
\label{sec:res_bino}
Light neutralino dark matter with a relic abundance of the order of the observed
value is realised for a $\chi^0_1$ with a sizable bino component. The bino
is a $SU(2)_L$ singlet with zero hypercharge.
As for a pure bino there are no interactions with electroweak gauge bosons nor
photons we can already expect that there will be essentially no long-range
potential interactions for the bino-like $\chi^0_1$ and hence no Sommerfeld
enhancements in $\chi^0_1\chi^0_1$ annihilations.
Yet it is interesting to confirm this expectation and to investigate the
relevance of Sommerfeld enhancements in possible co-annihilations
with (slightly) heavier neutralino and chargino states.
As an example for such a bino-like $\chi^0_1$ we chose to study the pMSSM
Snowmass model with ID $2178683$ that features wino-like NLSP states with masses
around $6\%$ heavier than the $\chi^0_1$ state:
$m_{\chi^0_1} = 488.8\,$GeV, $m_{\chi^0_2} = 516.0\,$GeV and
$m_{\chi^+_1} = 516.2\,$GeV.

In the calculation of the $\chi^0_1$ relic abundance we consider 
co-annihilation reactions among all $\chi^0/\chi^\pm$ two-particle states,
although only the two-particle annihilations between the states
$\chi^0_{1,2}, \chi^\pm_1$ 
are relevant since the higgsino-like states $\chi^0_{3,4}, \chi^\pm_2$ lie at the
$2\,$TeV scale and their abundances are strongly Boltzmann-suppressed
at $\chi^0_1$ freeze-out. The lightest sfermions are the $\tilde \tau_1$ and
$\tilde\nu_\tau$ with masses around $770\,$GeV and we neglect their effect in
the relic abundance.

\begin{figure}[t]
\begin{center}
\includegraphics[width=0.75\textwidth]{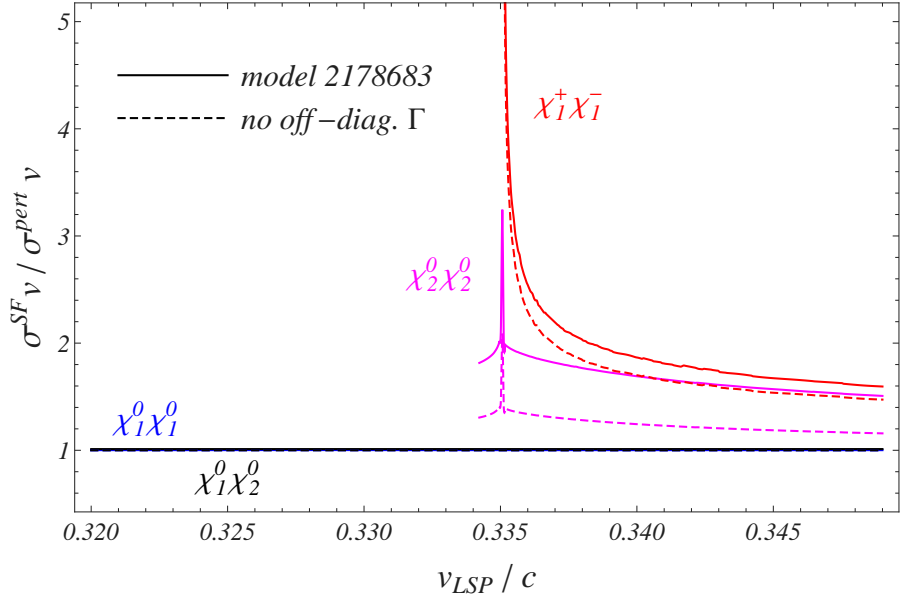}
\caption{ $(\sigma^\text{SF}v)/(\sigma^\text{pert}v)$ for the neutral-sector
  states in the light scenario (Snowmass model 2178683). Solid (dashed) curves
  show the enhancement for the case of properly included (wrongly neglected)
  off-diagonal annihilation rates.}
\label{fig:pMSSM_2178683_sigmavoff}
\end{center}
\end{figure}

Sommerfeld corrections on the co-annihilation cross sections
from all two-particle states built from $\chi^0_{1,2}$ and $\chi^\pm_1$
are determined exactly through the solution of the corresponding 
Schr\"odinger equations in each charge sector. The outcome for the
enhancement $(\sigma^\text{SF} v)/(\sigma^\text{pert} v)$
in the neutral sector,
which entails the two-particle states $\chi^0_1\chi^0_1$, $\chi^0_1\chi^0_2$,
$\chi^0_2\chi^0_2$ and $\chi^+_1\chi^-_1$,
is shown in Fig.~\ref{fig:pMSSM_2178683_sigmavoff}.
Solid (dashed) curves correspond to a calculation with (without) off-diagonal
annihilation rates in the Sommerfeld-enhanced reactions.
Due to the absence of interactions with the electroweak gauge bosons in case of
a pure-bino state, the $\chi^0_1$ of the pMSSM Snowmass model $217868$ also
experiences basically no long-range potential interactions and there is
essentially no coupling between the bino-like $\chi^0_1$ and the NLSP $\chi^0_2$.
As a consequence, both the absolute (perturbative as well as
Sommerfeld-enhanced)
$\chi^0_1\chi^0_1$ and $\chi^0_1\chi^0_2$ annihilation rates are
strongly suppressed and there is no enhancement in these reactions;
the ratio $(\sigma^\text{SF} v)/(\sigma^\text{pert} v)$ is equal to one
in both cases.
As it cannot be inferred from Fig.~\ref{fig:pMSSM_2178683_sigmavoff},
let us note in addition that the absolute $\chi^0_1\chi^0_1$
($\chi^0_1\chi^0_2$) annihilation cross section is suppressed with respect to the
dominant $\chi^0_2\chi^0_2$ and $\chi^+_1\chi^-_1$ rates by
four (two) orders of magnitude.

%---------------------------------------------------------------------------
%---------------------------------------------------------------------------

%---------------------------------------------------------------------------
\begin{figure}[t!]
\begin{center}
\includegraphics[width=0.75\textwidth]{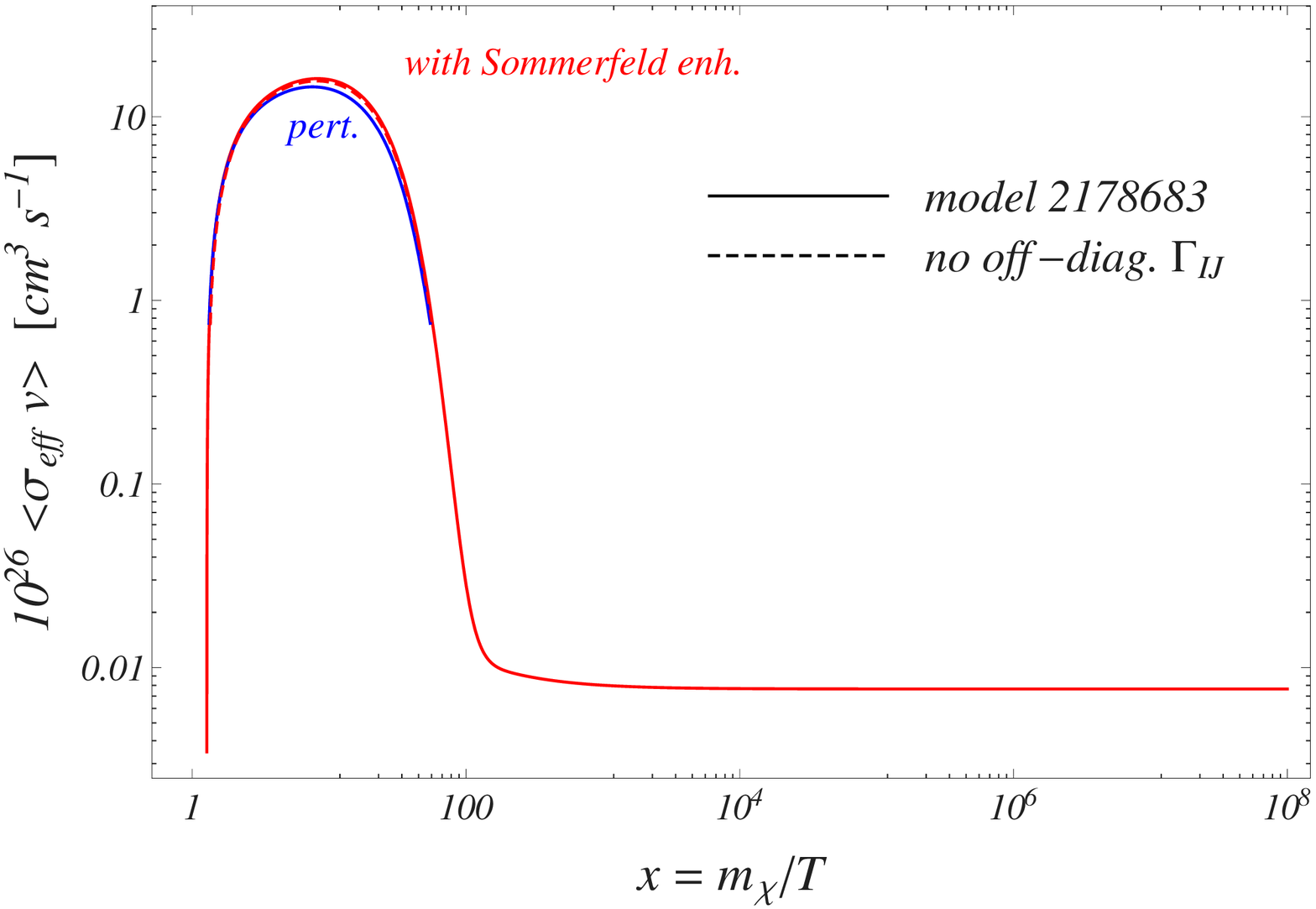}
\caption{The thermally averaged effective rate
         $\langle \sigma_\text{eff} v\rangle(x)$ within the pMSSM Snowmass model
         $2178683$ with Sommerfeld enhancements (upper red curve) and in the
         perturbative computation (lower blue curve).
         The result from disregarding off-diagonal rates in the
         Sommerfeld-enhanced processes is plotted by the dashed line.
         However the latter curve basically overlays with the upper (red)
         curve in this plot. This is because the Sommerfeld-enhanced
         $\langle \sigma_\text{eff} v\rangle(x)$ is dominated
         by the $\chi_2^0\chi_2^0$ and $\chi_1^+\chi_1^-$ rates (before
         $\chi^0_2$ and $\chi^\pm_1$ decoupling), and the
         effect of disregarding off-diagonals in the latter gives a correction
         of around $10\%$ only, see Fig.~\ref{fig:pMSSM_2178683_sigmavoff}.
       }
\label{fig:pMSSM_2178683_compareoff}
\end{center}
\end{figure}
%---------------------------------------------------------------------------
In the subsystem of the neutral wino-like two-particle channels
$\chi^0_2\chi^0_2$ and $\chi^+_1\chi^-_1$, the Sommerfeld enhancement due to
long-range potential interactions is effective,
see the corresponding curves in Fig.~\ref{fig:pMSSM_2178683_sigmavoff}. Note
that $\chi^0_2$ and $\chi^\pm_1$ co-annihilations should still be
relevant in the $\chi^0_1$ relic abundance calculation within the pMSSM scenario
$2178683$, as the threshold velocities for $\chi^0_2\chi^0_2$ and
$\chi^+_1\chi^-_1$ on-shell production are $v_{\chi^0_1}\, \lessim\, 0.34\,c$
and thus of the order of typical $\chi^0_1$ velocities during thermal
freeze-out. This scenario provides an example showing
that the criterion established before for including long-distance effects
among two-particle states with masses smaller than
$M_{\rm max}=2m_{\chi^0_1} + m_{\chi^0_1} v_{\rm max}^2$ 
and $v_{\rm max}=1/3$ should not be considered rigidly. Rather it has to be
reassessed according to the given MSSM spectra to avoid overlooking
interesting effects.
Consequently, in order to account for the wino-like subsystem formed by the
states $\chi^0_2\chi^0_2$ and $\chi^+_1\chi^-_1$ we have set
$v_{\rm max}=0.34$ in the light scenario. 
At very small velocities the enhancements in the $\chi^0_2\chi^0_2$ and
$\chi^+_1\chi^-_1$ channels show the characteristics discussed
already for the wino model in Sec.~\ref{sec:res_wino}:
In the $\chi^0_2\chi^0_2$ system we find resonances just below the
$\chi^+_1\chi^-_1$ threshold, smoothed out in
Fig.~\ref{fig:pMSSM_2178683_sigmavoff}.
The strength of the enhancement below and above this resonance region is a
combined effect of the (off-diagonal) Yukawa and the diagonal Coulomb potential 
interactions in the $\chi^+_1\chi^-_1$ system. In particular the enhancement is
finite below the $\chi^+_1\chi^-_1$ threshold.
To the contrary, the $\chi^+_1\chi^-_1$ channel shows the typical Coulomb-like
$1/v_{\chi^+_1}$ enhancement from the dominating photon-exchange potential at
velocities directly above its on-shell production threshold.
Opposed to the $\mathcal O(10^2)$ enhancements found in Sec.~\ref{sec:res_wino}, the
overall enhancements of the neutral wino-like two-particle channels here reach
factors of $\mathcal O(1)$ only. These less pronounced enhancements result
from the lower masses of the wino-like states,
since as  $m_{\chi_1^0}$ decreases the Yukawa potentials from electroweak gauge boson exchange
eventually become 
short-ranged as compared to the Bohr radius of the system proportional to
$(m_{\chi_1^0}\alpha_{\text{EW}})^{-1}$,
where $\alpha_\text{EW}=g_2^2/(4 \pi)$ and $g_2$ denotes the $SU(2)_L$ gauge
coupling.

%---------------------------------------------------------------------------
\begin{figure}[t]
\begin{center}
\includegraphics[width=0.75\textwidth]{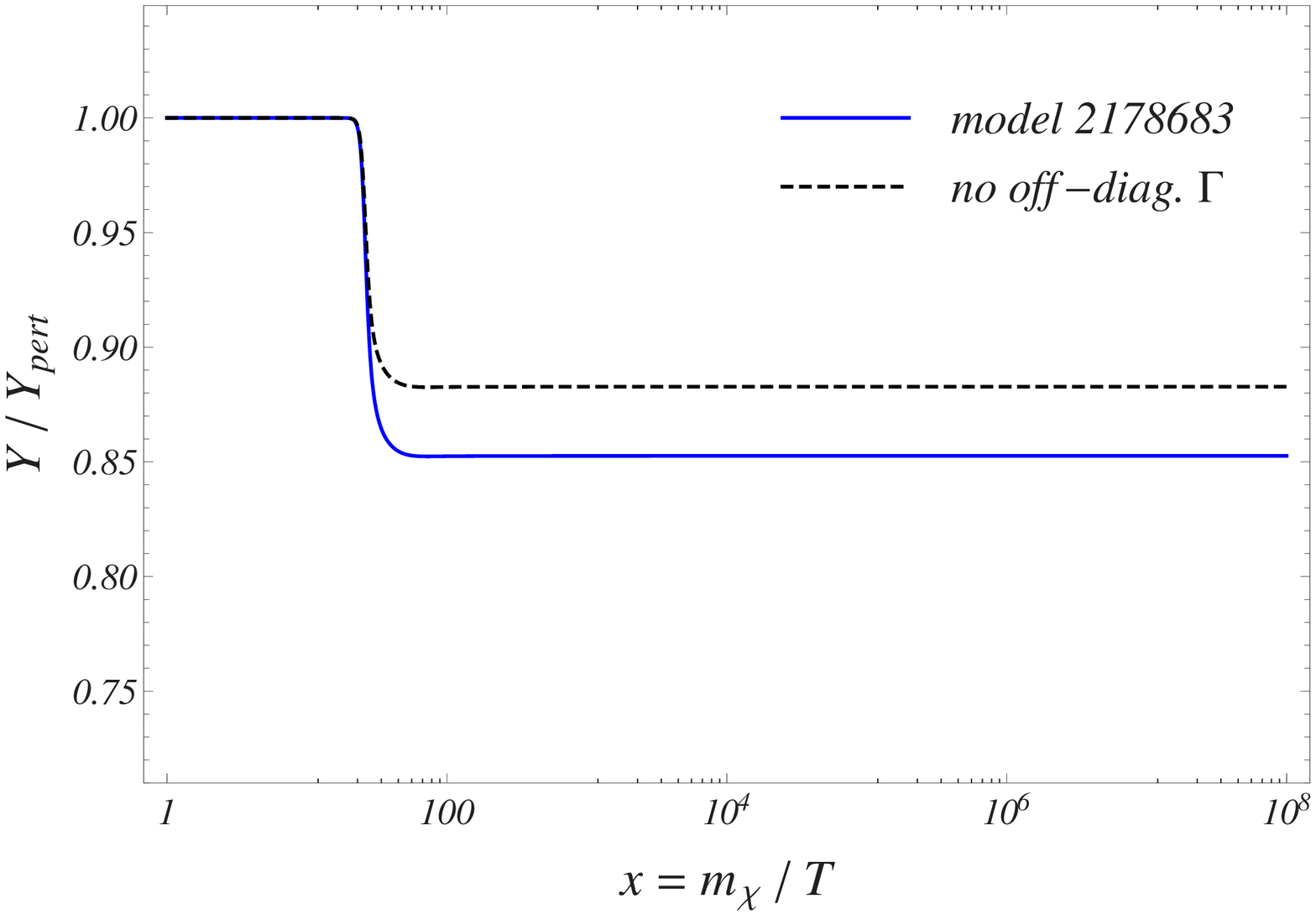}
\caption{The ratio of the yields $Y/Y_\text{pert}$, where
         $Y$ is calculated including the Sommerfeld enhancement on the
         annihilation rates and $Y_\text{pert}$ refers to the corresponding
         perturbative calculation. The solid (blue) line 
         includes off-diagonal rates while in the
         dashed (black) curve these have been neglected.
       }
\label{fig:pMSSM_2178683_Y}
\end{center}
\end{figure}
%---------------------------------------------------------------------------

Fig.~\ref{fig:pMSSM_2178683_compareoff} displays the effective
annihilation cross section $\langle\sigma_\text{eff} v \rangle(x)$. The
dominance of the wino-like $\chi^0_2, \chi^\pm_1$ particle annihilation rates
by more than three orders of magnitude before their decoupling 
near $x\sim 100$ is clearly visible. The Sommerfeld enhancement affects 
only the annihilation of the wino-like particles and therefore 
disappears for $x>100$. Although the Sommerfeld factors for these 
channels lead to ${\cal O}(1)$ enhancements of the cross sections 
above the threshold near $v_{\rm LSP}\sim 1/3$, similar in magnitude 
to the model with wino-like LSP for the same velocities, the thermal 
average over $v_{\rm LSP}$ dilutes the enhancement, since the cross section 
for the heavy channels vanishes below the threshold. 
Nevertheless, the small enhancement visible in 
Fig.~\ref{fig:pMSSM_2178683_compareoff} occurs precisely in the 
$x$ range most relevant for freeze-out. 
The effect of co-annihilations with the wino-like NLSP states
therefore leads to a reduction of the yield when taking into account 
Sommerfeld enhancements with respect to the perturbative case, 
as is shown in Fig.~\ref{fig:pMSSM_2178683_Y}. The relic density with
perturbative annihilation rates is found to be 
$\Omega^\text{pert} h^2 = 0.120$. 
There is a $\sim 15\%$ reduction of this result when considering the
Sommerfeld-enhanced rates, $\Omega^\text{SF} h^2 = 0.102$.
The latter sizable reduction of the relic density is attributed purely to the
co-annihilating heavier
wino states. Note that in the sector of wino-like states the potentials from
massive gauge boson and photon exchange are equally important for the Sommerfeld
enhancement, while in the $\chi^+_1\chi^-_1$ system the Coulomb potential 
dominates over the Yukawa potentials only for very
small velocities of the charginos.
Neglecting the perturbative correction from the heavier $\chi\chi$-states
not included in the Schr\"odinger equation leads essentially to no difference
(below per mil level) in the relic density, as the
heavy higgsino-like $\chi^0_{3,4}, \chi^\pm_2$ species lie at the scale of
around $2\,$TeV.
If no off-diagonals in the calculation of Sommerfeld-enhanced rates were
considered, the relic abundance would be overestimated by $3.5\%$.

\section{Higgsino-to-wino trajectory}
\label{sec:res_trajectory}
In case of the wino-like $\chi^0_1$ model of Sec.~\ref{sec:res_wino} we have
seen that the relic abundance including Sommerfeld enhancements on the
co-annihilation rates is reduced by about $40\%$ with respect to the result
calculated from tree-level annihilation rates. In contrast, the
model with higgsino-like $\chi^0_1$ in Sec.~\ref{sec:res_higgsino} shows a less
strong reduction, which is however still of the order of
$\Omega^\text{SF} h^2 / \Omega^\text{pert} h^2\approx 0.9$.
The difference in the reduction factor
$\Omega^\text{SF} h^2 / \Omega^\text{pert} h^2$
between the wino- and the higgsino-like $\chi^0_1$ model was explained by the
smaller Sommerfeld enhancements in the latter case due to larger
mass splittings between all co-annihilating particles and the fact that
the potential interactions happen to be generically weaker for higgsino-like
compared to the wino-like $\chi^0_1$ models.
In addition, we observed a Sommerfeld suppression effect in the single-charged
sector of the pure higgsino scenario as well as the higgsino-like Snowmass
model.
Departing from the scenarios with rather pure wino, higgsino or bino
$\chi^0_1$, we may ask ourselves about the features
of a model with $\chi^0_1$ LSP that contains both significant wino and
higgsino contributions.
It is worth to mention here that previous work in the literature focused on the
wino- or higgsino-like $\chi^0_1$ cases only, due to the lack of expression for
potentials and annihilation matrices for a generically composed $\chi^0_1$
state.
Our results allow for the first time to perform a rigorous study of Sommerfeld enhancements
in $\chi\chi$ pair-annihilations within models with mixed gaugino and
higgsino composition of the co-annihilating neutralinos and charginos.
We find it particular instructive to consider a series of models
in the MSSM parameter space that describes the transition from a model with
higgsino-like $\chi^0_1$ to a model with primarily wino-$\chi^0_1$.
In the following we will refer to this series of models as models on a
``higgsino-to-wino'' trajectory.
We are interested in the case of reductions of
$\Omega^\text{SF} h^2$ relative to $\Omega^\text{pert} h^2$ by $\gtrsim 10\%$
here and hence will
not consider a significant bino-admixture to the $\chi^0_1$; as we have seen in
Sec.~\ref{sec:res_bino} the bino-like $\chi^0_1$ itself does not experience any
Sommerfeld enhancement. In such a situation a reduction of
$\Omega^\text{SF} h^2$ can only arise due to co-annihilating particles with
Sommerfeld-enhanced rates, see for example the model discussed in
Sec.~\ref{sec:res_bino} with co-annihilating wino-like NLSPs.

In order to define the models for the higgsino-to-wino trajectory, 
we should note 
first that the proper choice of the two SUSY parameters $\mu$ and $M_2$
controls the higgsino and wino content of the mass eigenstate
$\chi^0_1$. In order to avoid a bino-admixture to the $\chi^0_1$ state we will
choose the parameter $M_1$, that controls the neutralinos'
bino-content, to be sufficiently larger than both $\mu$ and
$M_2$ throughout this section. Our setup excludes accidental
mass degeneracies of the MSSM sfermions with the $\chi^0_1$, which implies that
the actual parameters of the sfermion sector play a minor role in the choice of
adequate models on the trajectory. Let us recall that the sfermion sector is
irrelevant for Sommerfeld enhancements in our setup, as the latter are
caused by potential gauge boson and light Higgs
exchange between neutralino and chargino two-particle states prior to the
hard annihilation reactions. The sfermion sector parameters only affect
the precise value of the hard (tree-level) annihilation rates.
The sfermion -- basically the stop -- sector however controls the value of the 
Higgs $h^0$ mass and we will adjust its parameters such that
the experimental value for $m_{h^0}$ is reproduced within $2.5\%$ accuracy. Yet
matching the precise experimental Higgs mass value is in fact not important to
us here, as potential exchange from the $h^0$ gives always a sub-leading
contribution to the potentials compared to the effects from SM gauge boson
exchange.

\begin{figure}[t]
\begin{center}
\includegraphics[width=0.75\textwidth]{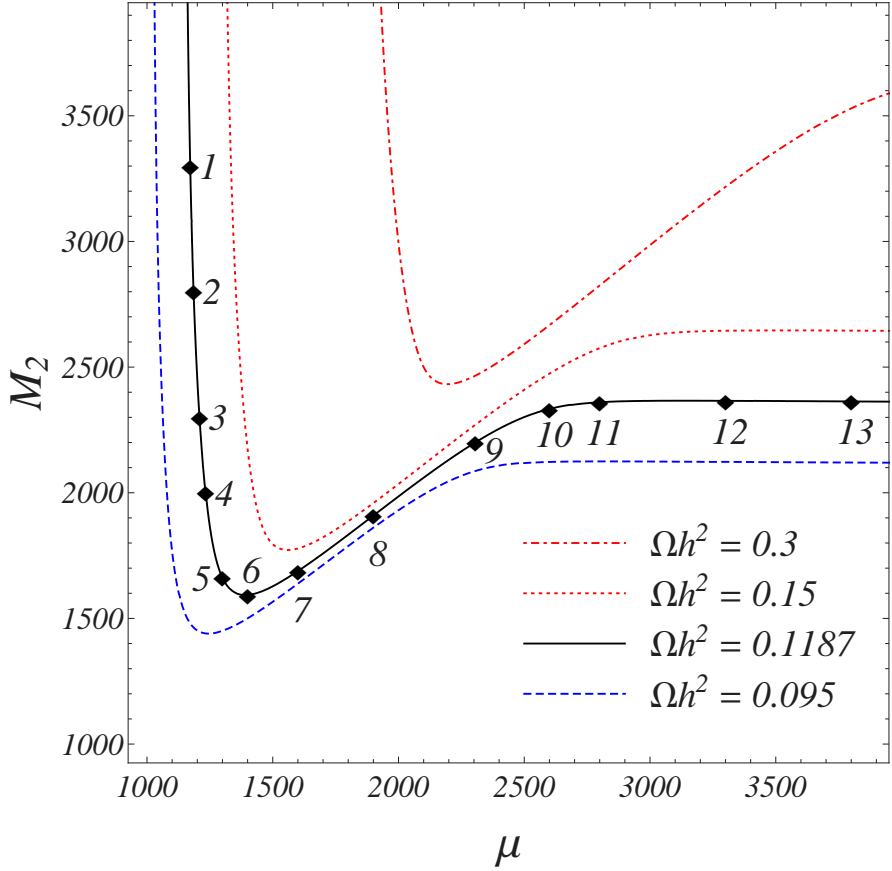}
\caption{ The $\mu-M_2$ plane with the $13$ models defining the higgsino-to-wino
         trajectory, indicated with diamonds.
         All trajectory models lie on the iso-contour for constant
         relic density $\Omega^\text{DS} h^2 = 0.1187$ calculated with DarkSUSY.
         As reference we also show the iso-contours of constant relic densities
         $\Omega^\text{DS} h^2 = 0.095$ (lowermost contour-line)
         $0.15$ and $0.3$ (uppermost iso-contour).
       }
\label{fig:res_muM2plane}
\end{center}
\end{figure}

In order to generate MSSM scenarios on a higgsino-to-wino trajectory we hence
make the following choice for MSSM input parameters in the spectrum generation:
\begin{itemize}
 \item fix a common sfermion mass scale of $9\,$TeV,
 \item set the trilinear couplings to $A_t = A_b = 9\,$TeV,
 \item fix $m_{A^0} = 500\,$ GeV and
 \item choose $\tan\beta = 15$.
\end{itemize}
All other trilinear couplings are assumed to vanish.
The gluino mass parameter $M_3$ is fixed by
$M_3 = \alpha_s/(\sin(\theta_\text{w})\,\alpha_e) \, M_2$, but this choice
is completely irrelevant to our discussion.
To avoid a significant
bino-admixture to the $\chi^0_1$ we further restrict to models
with $M_1 = 10\,M_2$.
This leaves us with yet-to-choose parameter pairs in the $\mu - M_2$ plane.
We require that the trajectory models allow for an explanation of the observed
cosmic cold dark matter in terms of the neutralino relic abundance without
including radiative corrections: in order to do so
we employ the program DarkSUSY \cite{Gondolo:2004sc}
and identify $(\mu,M_2)$ pairs such that the DarkSUSY calculated relic density
$\Omega^\text{DS}h^2$ matches the most accurate determination obtained
from the combination of PLANCK, WMAP, BAO and high resolution CMB data,
$\Omega_\text{cdm} h^2 = 0.1187\pm0.0017$~\cite{Ade:2013zuv}\footnote{Note 
that the DarkSUSY collaboration claims an error of $5\%$ on the relic densities calculated from their code.}.
In such a way we define $13$ models on the higgsino-to-wino trajectory.
The position of these models in the $\mu-M_2$ plane is shown in
Fig.~\ref{fig:res_muM2plane}.
For each of the $13$ models, given the pairs $(\mu, M_2)$ as well as the
remaining  input parameters defined above, we run our code and determine the
corresponding relic densities including and neglecting Sommerfeld effects.
The comparison between our perturbative results $\Omega^\text{pert} h^2$ with the
corresponding DarkSUSY expressions $\Omega^\text{DS} h^2$ provides a
cross-check of our perturbative calculation.
%---------------------------------------------------------------------------
%---------------------------------------------------------------------------

There is one important point to note concerning the MSSM spectrum generation
from the SUSY input parameters. The DarkSUSY spectrum calculated from the inputs
refers to tree-level $\overline{\text{DR}}$-parameters. It is well-known 
that the mass splitting between a wino-like neutralino and its chargino
partner is dominated by radiative corrections; the leading one-loop
contribution to the splitting is of $\mathcal O(160\,$MeV$)$
and dominates over the $\mathcal O(1\,$MeV$)$ tree-level contribution.
Both for the calculation of the Sommerfeld enhancements and in the
determination of the relic abundance including co-annihilations a precise
knowledge of the mass splitting between the $\chi^0_1$ LSP and the NLSP
particles is crucial and in a rigorous analysis we should therefore consider
the spectra determined with one-loop accuracy.
To this end we have been provided by one-loop on-shell renormalised
SUSY spectra for all $13$ models on the trajectory by a member of the
collaboration \cite{Bharucha:2012re,Bharucha:2012nx}.
The values of the input parameters $\mu, M_2, \ldots$ are the same as
for the corresponding calculation within DarkSUSY with the difference that for
the one-loop on-shell spectrum
generation these inputs are considered as on-shell parameters and no
renormalisation group running of the mass parameters is performed. Hence there
are small differences in the values for the masses and mixing-matrix entries
between the spectra that we use in our code and the corresponding DarkSUSY
spectra. In particular the mass splittings between the $\chi^0_1$ LSP and the
NLSPs obtained from the on-shell masses renormalised at one-loop can be significantly
different from the splittings derived using tree-level
$\overline{\text{DR}}$-parameters.
There exist different renormalisation schemes for on-shell renormalisation in
the neutralino/chargino sector
\cite{Heinemeyer:2011gk, Bharucha:2012re, Bharucha:2012nx, Fritzsche:2013fta}:
for all trajectory models apart from model $8$
the on-shell renormalisation has been performed requiring that the values of the two
chargino masses as well as the heaviest (in all our models bino-like) neutralino
mass at one-loop are given by their tree-level values (``CCN-scheme'').
Such a scheme works well as long as the two charginos are rather pure
wino- and higgsino-like states.
As soon as the charginos are (strongly) mixed wino-higgsino states - as in case
of our model $8$, where the input parameters $\mu$ and $M_2$ happen to be
very close to each other - a more suitable scheme is obtained when only one chargino,
one lighter neutralino and the heaviest bino-like neutralino mass are fixed to
their tree-level value (``CNN scheme'').

\begin{figure}[t]
\begin{center}
\includegraphics[width=0.75\textwidth]{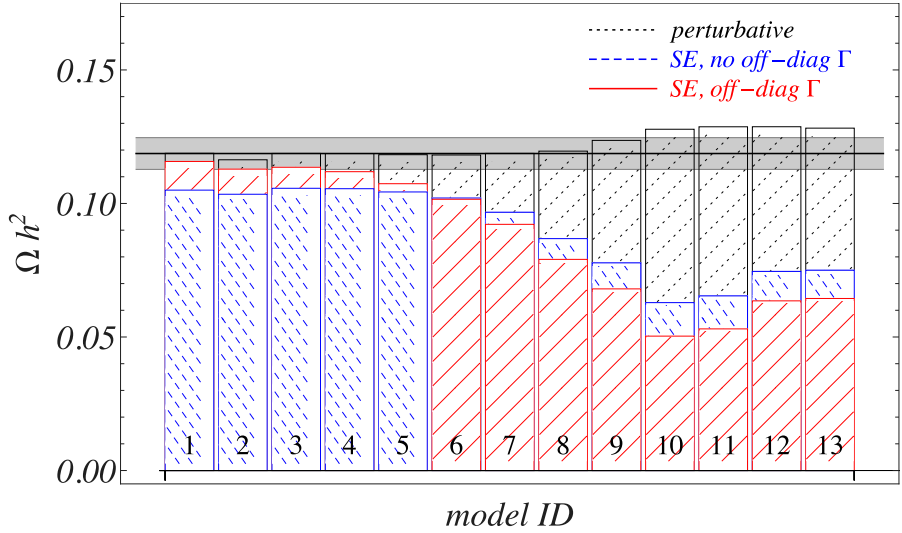}
\caption{ Relic densities $\Omega h^2$ for models $1-13$ on
          the higgsino-to-wino trajectory calculated with our code.
          The charts with dotted (black) hatching are the perturbative results
          $\Omega^\text{pert} h^2$. Bars with dashed (blue) and solid-line (red)
          hatching refer to a calculation with Sommerfeld-enhanced
          cross sections neglecting and properly including off-diagonal rates,
          respectively.
          The grey shaded band comprises $\Omega h^2$ values within
          $5\%$ around the mean experimental value
          $\Omega_\text{cdm} h^2 = 0.1187$~\cite{Ade:2013zuv}.
          The latter value is indicated by the black horizontal line and
          agrees with the DarkSUSY result for all $13$ MSSM models on the
          trajectory.
       }
\label{fig:trajectory}
\end{center}
\end{figure}

%-----------------------------------------------------------
\begin{table}[t]
\centering
\begin{tabular}{|c|c|c|c|c|c|c|c|}
\hline
   ID
 & $\mu/$GeV
 & $M_2/$GeV
 & $m_{\chi^0_1}/$GeV
 & $\delta m_{\chi^+_1}/$GeV
 & $\vert Z_{N\,21} \vert^2$
 & $\Omega^\text{SF} h^2$
 & $\frac{\Omega^\text{SF} h^2}{\Omega^\text{pert} h^2}$
\\ &&&&&&&\vspace{-1.35em}
\\
\hline\hline
   1
 & 1171.925
 & 3300.000
 & 1169.957
 & 0.876
 & 0.001
 & 0.1157
 & 0.974
\\
   2
 & 1185.224
 & 2800.000
 & 1169.427
 & 0.958
 & 0.001
 & 0.1129
 & 0.970
\\
   3
 & 1208.699
 & 2300.000
 & 1205.096
 & 1.057
 & 0.003
 & 0.1136
 & 0.956
\\
   4
 & 1233.685
 & 2000.000
 & 1228.674
 & 1.129
 & 0.006
 & 0.1119
 & 0.943
\\
   5
 & 1300.000
 & 1661.705
 & 1289.890
 & 1.203
 & 0.026
 & 0.1074
 & 0.908
\\
   6
 & 1400.000
 & 1593.100
 & 1382.390
 & 1.153
 & 0.076
 & 0.1016
 & 0.860
\\
   7
 & 1600.000
 & 1688.240
 & 1569.117
 & 0.971
 & 0.203
 & 0.0922
 & 0.776
\\
   8
 & 1900.000
 & 1909.355
 & 1844.126
 & 0.601
 & 0.458
 & 0.0791
 & 0.661
\\
   9
 & 2304.666
 & 2200.000
 & 2172.690
 & 0.266
 & 0.826
 & 0.0680
 & 0.550
\\
   10
 & 2600.000
 & 2333.7034
 & 2320.986
 & 0.183
 & 0.955
 & 0.0503
 & 0.394
\\
   11
 & 2800.000
 & 2360.2715
 & 2352.475
 & 0.166
 & 0.982
 & 0.0530
 & 0.412
\\
   12
 & 3300.000
 & 2365.830
 & 2362.264
 & 0.158
 & 0.996
 & 0.0635
 & 0.494
\\
   13
 & 3800.000
 & 2363.500
 & 2361.254
 & 0.157
 & 0.998
 & 0.0644
 & 0.503
\\
\hline
\end{tabular}
\caption{Information on the models on the higgsino-to-wino trajectory.
The first column is the model ID while the second and third column contain
the input parameter values for $\mu$ and $M_2$. The one-loop on-shell renormalised
$\chi^0_1$ LSP mass is given in the fourth column and we provide
the one-loop mass splitting to the lighter chargino,
$\delta m_{\chi^+_1} = m_{\chi^+_1} - m_{\chi^0_1}$ in the fifth column.
The $\chi^\pm_1$ are the NLSP states in all models considered here.
In the sixth column the wino fraction, $\vert Z_{N\,21}\vert^2$, of the $\chi^0_1$
is specified.
The second-to-last and the last columns give the relic
density including Sommerfeld-enhanced cross sections as
well as the suppression factor of the $\Omega^\text{SF} h^2$ with respect to the
perturbative result $\Omega^\text{pert} h^2$.
The results including the Sommerfeld enhancements involve corrections from
heavier $\chi\chi$-pairs in the last potential loop.
}
\label{tab:res_trajectory}
\end{table}
%-----------------------------------------------------------
For each of the $13$ models on the trajectory we list the input parameters
$\mu$ and $M_2$ in Tab.~\ref{tab:res_trajectory}, together with the one-loop
renormalised LSP mass $m_{\chi^0_1}$ as well as the one-loop on-shell mass splitting
$\delta m_{\chi^+_1} = m_{\chi^+_1} - m_{\chi^0_1}$. The $\chi^\pm_1$ is the NLSP in
all models considered in this section. As additional information we
give the $\chi^0_1$'s wino fraction $\vert Z_{N\,21}\vert^2$ and collect the
results for $\Omega^\text{SF} h^2$ including Sommerfeld effects as well as for
the suppression $\Omega^\text{SF} h^2 / \Omega^\text{pert} h^2$ of the former
relic density with respect to the perturbative result.
Both $\Omega^\text{SF} h^2$ and $\Omega^\text{pert} h^2$ are calculated from
our programs, and the latter shows small deviations of the order of a
few percent from the DarkSUSY value $\Omega^\text{DS} h^2 = 0.1187$.
As can be read off Tab.~\ref{tab:res_trajectory} we can categorise the models
on the trajectory to feature either a higgsino-like $\chi^0_1$ with
wino fraction below $10\%$ but a higgsino fraction
$\vert Z_{N\, 31}\vert^2 + \vert Z_{N\, 41}\vert^2$ above $0.9$ (models $1-6$),
a mixed wino-higgsino $\chi^0_1$ where both the wino and the higgsino fraction
lie within $0.1 - 0.9$ (models $7-9$) or a predominantly wino-like $\chi^0_1$
with wino fraction above $0.9$ (models $10-13$).
For all models we collect the relic density results $\Omega^\text{pert} h^2$ and
$\Omega^\text{SF} h^2$ in Fig.\ref{fig:trajectory}.
The bars with dotted (black) hatching indicate $\Omega^\text{pert} h^2$. Bars
with solid-line (red) and dashed (blue) hatching give the corresponding
results including Sommerfeld enhancements with and without
off-diagonal rates, respectively.
In particular for the higgsino-like models $1 - 6$ but also for models
$7 - 9$ our relic densities $\Omega^\text{pert} h^2$ agree very well with the
relic density $\Omega^\text{DS}h^2 = 0.1187$ calculated with DarkSUSY for the
same set of input parameters.
The latter relic density value is indicated by the black horizontal line and
the grey horizontal band comprises all values deviating at most by $5\%$ from
the $\Omega^\text{DS} h^2$ value.
For the wino-like models our relic density results
deviate by $\lessim \,8\%$ from the corresponding DarkSUSY value.

%---------------------------------------------------------------------------
%---------------------------------------------------------------------------

Let us discuss the characteristics of the models in the three different classes
corresponding to their wino and higgsino admixture in turn.
The models $1 - 6$, with predominant higgsino composition,
resemble the higgsino model of Sec.~\ref{sec:res_higgsino}. This applies also to
the corresponding shapes of the Sommerfeld-enhanced rates $\sigma^\text{SF}v$,
$\langle\sigma_\text{eff}v\rangle$, as well as to the yields
$Y / Y^\text{pert}$, that we do not show here.
The reduction in the relic density when taking the
Sommerfeld effect into account ranges from $3\%$ to $14\%$ for
trajectory models $1 - 6$.
Models $1 - 3$, with a $3\%$ to $4\%$ reduction are close to a pure-higgsino
limit behaviour,
whereas models $4 - 6$ yield a similar outcome as for the
Sec.~\ref{sec:res_higgsino} higgsino-like $\chi^0_1$ Snowmass model.
The potential interactions among all 
two-particle states built from the higgsino-like particles
$\chi^0_{1,2}, \chi^\pm_1$ have been accounted for exactly by solving the corresponding multi-state
Schr\"odinger equation in models $1 - 6$. This is in agreement with the
criterion introduced in Sec.~\ref{sec:res_wino} that considers the long-distance effects among
all $\chi\chi$-states with mass smaller than
$M_\text{max} = 2\,m_{\chi^0_1} + m_{\chi^0_1} v_\text{max}^2$, where
$v_\text{max} = 1/3$ is of the order of the
$\chi^0_1$'s mean-velocity during freeze-out. 
Heavier $\chi\chi$ channels enter the calculation through the 
perturbative corrections to the annihilation rates of the lighter channels
treated exactly, and their tree-level co-annihilation rates are
also included in the calculation of the $\chi^0_1$ relic density,
as done in the previous sections.
The effect of neglecting off-diagonal annihilation rates in the determination
of $\Omega^\text{SF} h^2$ yields an error of about $9\%$ to $3\%$ for models
$1 - 5$, underestimating the true result.
In case of model $6$ the $\Omega^\text{SF}h^2$ results obtained when
neglecting or correctly including off-diagonal annihilation rates happen to
agree.
This can be understood from the Sommerfeld suppressions in the two
single-charged sectors that arise when correctly accounting for off-diagonal
annihilation rates and that can lead to a partial compensation of enhancements
encountered in the neutral sector.
While there is no suppression effect if off-diagonal annihilation rates are
neglected, also the Sommerfeld enhancements in the charge-neutral sector are
milder in that case, see for instance
Fig.~\ref{fig:pMSSM_1627006_sigmavoff_Coulomb}.
Relic density results with and without off-diagonal annihilation rates can
therefore accidentally agree, as it happens for model $6$.
If corrections from heavier states in the last potential loop were not
included in the calculation of the relic abundance, the corresponding result
would be larger by $2\%$ for model $1$ to $6\%$ for model $6$ as
compared to the $\Omega^\text{SF} h^2$ values quoted in
Tab.~\ref{tab:res_trajectory}.
As  expected, the latter effect gains importance
as the mass splitting of the heavier states to the
higgsino-like $\chi^0_{1,2}$ and $\chi^\pm_1$ becomes smaller; while the wino-like states
$\chi^0_3, \chi^\pm_2$ in model $1$ are rather heavy $(m\sim 3.3\,$TeV$)$, these 
states have a mass of about $1.6\,$TeV in case of model $6$.

For models $7-9$ with mixed wino-higgsino $\chi^0_1$, where the
wino content increases with higher model ID,
Fig.~\ref{fig:trajectory} shows a reduction of $\Omega^\text{SF} h^2$ the
larger the wino admixture of the $\chi^0_1$. 
The ratio $\Omega^\text{SF} h^2 / \Omega^\text{pert} h^2$ ranges from
$\sim 0.78$ for model $7$ over $\sim 0.66$ for model $8$ and gives
$\sim 0.55$ in case of model $9$.
In the region of mixed wino-higgsino $\chi^0_1$, where the masses of the states
$\chi^0_{1,2,3}, \chi^\pm_{1,2}$ lie close to each other, more two-particle
states have been considered exactly in the multi-state Schr\"odinger equation.
Precisely, the set of
neutral $\chi\chi$-states considered in the Schr\"odinger equations
for model 7 comprises the seven states $\chi^0_1\chi^0_1, \chi^+_1\chi^-_1,
\chi^0_1\chi^0_2, \chi^0_2\chi^0_2, \chi^0_1\chi^0_3, \chi^\pm_1\chi^\mp_2$,
while for model $8$ the state $\chi^0_2\chi^0_3$ is
included in addition, and for model $9$ only the  six states
$\chi^0_1\chi^0_1, \chi^+_1\chi^-_1, \chi^0_1\chi^0_2, \chi^0_1\chi^0_3,
\chi^\pm_1\chi^\mp_2$ are treated exactly in the neutral sector.
While in the three models $7-9$ (particularly in the neutral sector), the mutual
interaction among a large number of channels is solved through the Schr\"odinger
equations, it is mainly the larger wino fraction of the $\chi^0_1$ that
controls the increasing relevance of the Sommerfeld enhancements on the final
relic abundance.
While the wino fraction of the $\chi^0_1$ in model $7$ is $20\%$ it becomes
$46\%$ for model $8$ and finally reaches $83\%$ in case of model $9$. The larger
wino admixture of both the $\chi^0_1$ and $\chi^\pm_1$ states also
manifests itself in the decreasing mass splitting $\delta m_{\chi^+_1}$ between
these two states, ranging from $0.971\,$GeV (model $7$) over $0.601\,$GeV
(model $8$) to only $0.266\,$GeV (model $9$).
A larger wino component of the $\chi^0_1$ implies stronger potential
interactions between the co-annihilating channels, in
particular the $\chi^0_1\chi^0_1$ and $\chi^+_1\chi^-_1$, where the latter is
composed of $\chi^\pm_1$ states with similar wino fraction as the $\chi^0_1$.
The stronger potential interactions finally lead to a more pronounced
Sommerfeld enhancement effect for models with larger wino admixture to the
$\chi^0_1$ state.
Neglecting off-diagonal annihilation rates would lead to a result enhanced by
$5\%$ (model $7$), $10\%$ (model $8$) and $14\%$ (model $9$) with respect to
the actual $\Omega^\text{SF} h^2$ values given in Tab.~\ref{tab:res_trajectory}.
On the other hand, corrections to the Sommerfeld-enhanced rates 
from heavy $\chi\chi$-states in the last potential loop reduce
the final relic abundances
$\Omega^\text{SF} h^2$ for models $7 - 9$ by around $2 - 4\%$.
The latter reduction is not as large as for model $6$, despite the 
fact that the mass differences in
models $7 - 9$ are smaller. This is simply because there are less heavy
channels contributing perturbatively now, as more $\chi\chi$-states
have been considered exactly in the   Schr\"odinger equation.

Finally let us consider the subclass of wino-like $\chi^0_1$ models with
IDs $10 - 13$. Here we account for Sommerfeld effects on the
annihilation rates for $\chi\chi$-states built from the
wino-like $\chi^0_1$ and $\chi^\pm_1$ particles.
The  Schr\"odinger equations in the neutral sector for models $10 - 13$ hence contain
the two states $\chi^0_1\chi^0_1$ and $\chi^+_1\chi^-_1$ only.
The models can be further subdivided into two groups with
different impact of Sommerfeld enhancements: in case of models $10$ and $11$,
$\Omega^\text{SF} h^2$ is significantly reduced by around $60\%$
with respect to the result from a perturbative calculation. This happens
to be the strongest reduction we find along the trajectory.
The reason for the especially pronounced
Sommerfeld-enhanced annihilation rates in case of models $10$ and
$11$ can be attributed to the presence of a so called zero-energy resonance~\cite{Hisano:2004ds} in
the $\chi^0_1\chi^0_1$ annihilation channel:
as already discussed, for velocities well below the $\chi^+_1\chi^-_1$ threshold
the enhancement in the $\chi^0_1\chi^0_1$ system is controlled by the Yukawa
potential due to electroweak $W$-exchange. As any
short-ranged potential, a Yukawa-potential features a finite number of bound
states. 
By varying the potential's strength and range it is possible to arrange for the
presence of a bound state with (almost) zero binding energy~\cite{Hisano:2004ds}
(see also~\cite{Slatyer:2009vg}).
In the presence of such a (loosely) bound state, the scattering cross section
for incoming particles with very low velocities is strongly enhanced.
This effect leads to
$\mathcal O(10^4)$ enhancements in the $\chi^0_1\chi^0_1$ channel for
velocities below the $\chi^+_1\chi^-_1$ threshold and eventually translates
into the pronounced reduction of about $60\%$ of the relic density.
If off-diagonal annihilation rates were not taken into account, the
$\Omega^\text{SF} h^2$ result would be larger by about $25\%$ (model $10$) and
$23\%$ (model $11$), thus representing a rather large effect for both models:
Off-diagonal annihilation rates are particularly important if the
corresponding off-diagonal potential interactions are sufficiently
strong.
In wino-like $\chi^0_1$ models, the only sector with relevant off-diagonal
potential interactions is given by the two neutral states $\chi^0_1\chi^0_1$ and
$\chi^+_1\chi^-_1$ in a ${}^1S_0$ wave configuration.\footnote{
To a lesser extent, as it constitutes higher partial waves, also the
${}^3P_{\cal J}$ configurations are important.}
For models $10$ and $11$, where the neutral $\chi^0_1\chi^0_1$ channel
experiences particularly large enhancements due to the presence of a (loosely) bound
state resonance related to the off-diagonal 
$W$-exchange potential, also the impact of off-diagonal annihilation rates is
therefore found to be significant. Regarding the
corrections from heavier $\chi\chi$-states treated perturbatively in the
last potential loop, they are rather mild: $\Omega^\text{SF} h^2$ would be
smaller by around $3\%$ without this effect.
Compared to model $6$, where we found a corresponding $6\%$ reduction in
$\Omega^\text{SF} h^2$, this suggests that the effect from heavier
$\chi\chi$-states in the last potential loop is most significant if these states
are built from wino-like particles. The latter have in overall stronger (off-)
diagonal annihilation rates compared to higgsino-like states with similar mass.
Let us recall that the effect from heavier $\chi\chi$-states in the last
potential loop was at the per mil level in case of the pMSSM scenarios in
Secs.~\ref{sec:res_wino} and \ref{sec:res_bino} and around $1\%$ for the
higgsino-like scenario in Sec.~\ref{sec:res_higgsino}, because heavier
states were essentially decoupled in these models, opposed to the case for
the models on the higgsino-to-wino trajectory.

At last, for models $12$ and $13$ we find a reduction of $\Omega^\text{SF}h^2$
relative to $\Omega^\text{pert} h^2$ of roughly $50\%$ in both cases.
This is still larger than the $40\%$ reduction arising in case of
the wino-like $\chi^0_1$ pMSSM Snowmass model discussed in
Sec.~\ref{sec:res_wino}.
To explain this effect note first that
although the input value $\mu$ differs for models $12$ and $13$, this does not
affect the parameters of the corresponding wino-like sectors.
The masses of both $\chi^0_1$ and $\chi^\pm_1$ as well as their
wino fractions are essentially the same in model $12$ and $13$, see
Tab.~\ref{tab:res_trajectory}. We can hence expect that the results for the
$\chi^0_1$ relic abundance calculation are very similar for both models.
The presence of a zero-energy
resonance in the $\chi^0_1\chi^0_1$ annihilation channel is still noticeable
for models $12, 13$ -- although it is less pronounced, as increasing
the $\chi^0_1$ mass moves us away from the exact resonance region.
To conclude with the comparison to the wino-like $\chi^0_1$ pMSSM Snowmass
model in Sec.~\ref{sec:res_wino}, recall that the mass of the wino-like
$\chi^0_1$ there was $m_{\chi^0_1} = 1650.664\,$GeV; in that case
the Yukawa potential does not exhibit (almost) zero-energy bound states.
Consequently no additional strong resonant enhancement takes place, such that
in comparison to the wino-like models on the trajectory
the Sommerfeld effect on the relic density is less prominent in
Sec.~\ref{sec:res_wino}, though still around $40\%$.
Finally the calculated relic density $\Omega^\text{SF} h^2$ for both
models $12$ and $13$ is increased by $17\%$ and $16\%$, respectively, if
off-diagonal annihilations are neglected.
Not including the one-loop effects from heavy $\chi\chi$-states increases
the corresponding results for $\Omega^\text{SF} h^2$ in
Tab.~\ref{tab:res_trajectory} by $2\%$ in both cases.

\section{Mixed wino-higgsino $\chi^0_1$}
\label{sec:res_winohiggsino}
As our framework allows for the first time to investigate Sommerfeld
enhancements of $\chi\chi$ co-annihilations in scenarios with a $\chi^0_1$ in an
arbitrary wino-higgsino admixture, let us discuss here in  more detail
the mixed wino-higgsino $\chi^0_1$ trajectory model with ID $8$ considered in
the previous section.
Recall from section \ref{sec:res_trajectory} that the neutral sector
of the Schr\"odinger equation for this model is composed of the
eight states $\chi^0_1\chi^0_1$, $\chi^+_1\chi^-_1$,
$\chi^0_1\chi^0_2$, $\chi^0_2\chi^0_2$, $\chi^0_1\chi^0_3$, $\chi^\pm_1\chi^\mp_2$,
$\chi^0_2\chi^0_3$.

%---------------------------------------------------------------------------
\begin{figure}[t]
\begin{center}
\includegraphics[width=0.75\textwidth]{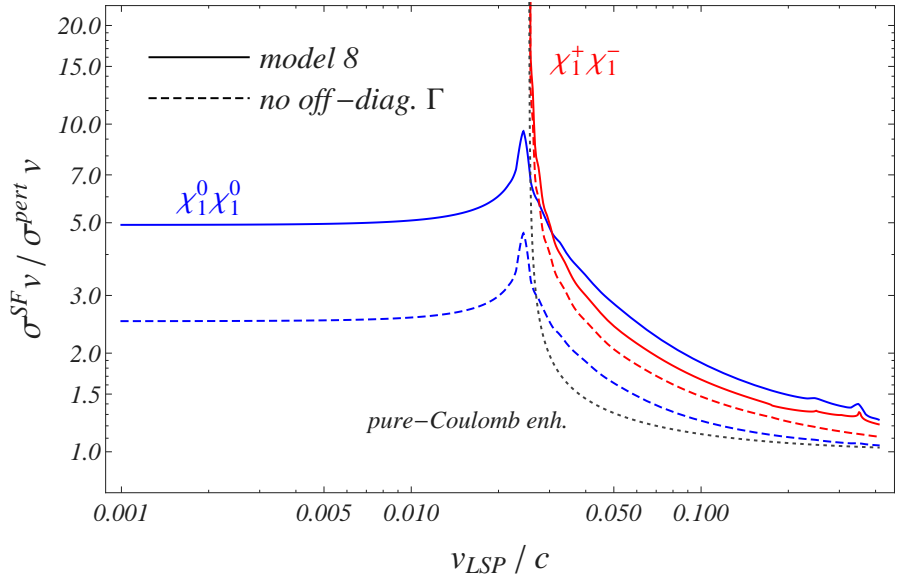}
\caption{ Enhancements $(\sigma^\text{SF} v)/(\sigma^\text{pert} v)$ in the
          two neutral channels $\chi^0_1\chi^0_1$ and $\chi^+_1\chi^-_1$ of
          model $8$ of the  wino-to-higgsino trajectory
          discussed in Sec.~\ref{sec:res_trajectory}.
          Solid (dashed) curves refer to the results with (without) 
          off-diagonal annihilation rates included.
       }
\label{fig:trajectory08_sigmav}
\end{center}
\end{figure}
%---------------------------------------------------------------------------

Fig.~\ref{fig:trajectory08_sigmav}
shows the enhancements $(\sigma^\text{SF} v)/(\sigma^\text{pert} v)$ in the
two neutral channels $\chi^0_1\chi^0_1$ and  $\chi^+_1\chi^-_1$ with (solid
lines) and without (dashed lines) off-diagonal annihilation rates.
The characteristic velocity-independent enhancement from the
$W$-exchange Yukawa potential in the low velocity regime of the $\chi^0_1\chi^0_1$ channel
is visible, as well as the Coulomb-type $1/v_{\chi^+_1}$ enhancement for the
$\chi^+_1\chi^-_1$ system at very low velocities.
Long-range potential interactions, although stronger than
in case of higgsino-like $\chi^0_1$ models are still weaker than in case
of a wino-like set of states $\chi^0_1, \chi^\pm_1$; as a consequence
enhancement factors of $\mathcal O(1-10)$ result.
We do not show $(\sigma^\text{SF} v)/(\sigma^\text{pert} v)$ for the remaining six
neutral two-particle states in Fig.~\ref{fig:trajectory08_sigmav},
but the resonance regions below their corresponding on-shell production
thresholds can be seen as small enhancements in the $\chi^0_1\chi^0_1$ and
$\chi^+_1\chi^-_1$ channels.
The threshold for $\chi^0_1\chi^0_2$ production opens at
$v_\text{LSP}/c \simeq 0.18$ but is hardly visible in the curves for channels
$\chi^0_1\chi^0_1$ and $\chi^+_1\chi^-_1$ in Fig.~\ref{fig:trajectory08_sigmav}.
We can notice a broader (smoothed-out) resonance region around
$v_\text{LSP}/c\simeq 0.25$, which comprises the thresholds for the four channels
$\chi^0_2\chi^0_2, \chi^0_1\chi^0_3$ and $\chi^\pm_1\chi^\mp_2$.
Finally, the
$\chi^0_2\chi^0_3$ threshold shows up at $v_\text{LSP}/c\simeq 0.30$.
The enhancements for these channels, not shown in
Fig.~\ref{fig:trajectory08_sigmav}, are somewhat smaller than for the cases of
$\chi^0_1\chi^0_1$ and $\chi^+_1\chi^-_1$.
Eventually, at $v_\text{LSP}/c \simeq 0.35$ the threshold for on-shell
production of the $\chi^0_3\chi^0_3$ state is visible in the $\chi^+_1\chi^-_1$
channel. The $\chi^0_3\chi^0_3$ state is among the heavy states considered perturbatively
in the last potential loop for the calculation of the annihilation rates of the channels
treated exactly in the neutral sector.

Note that apart from the bino-like $\chi^0_4$ state, which is 
very heavy ($m_{\chi^0_4}\sim19$~TeV) and -- being bino-like -- couples
very weakly to  the gauge bosons and the other $\chi^0/\chi^\pm$ species,
all $\chi$ states in the neutralino/chargino sector are relevant in
co-annihilation reactions for the $\chi^0_1$ relic abundance calculation
of model $8$.

%---------------------------------------------------------------------------
\begin{figure}[p]
\begin{center}
\includegraphics[width=0.75\textwidth]{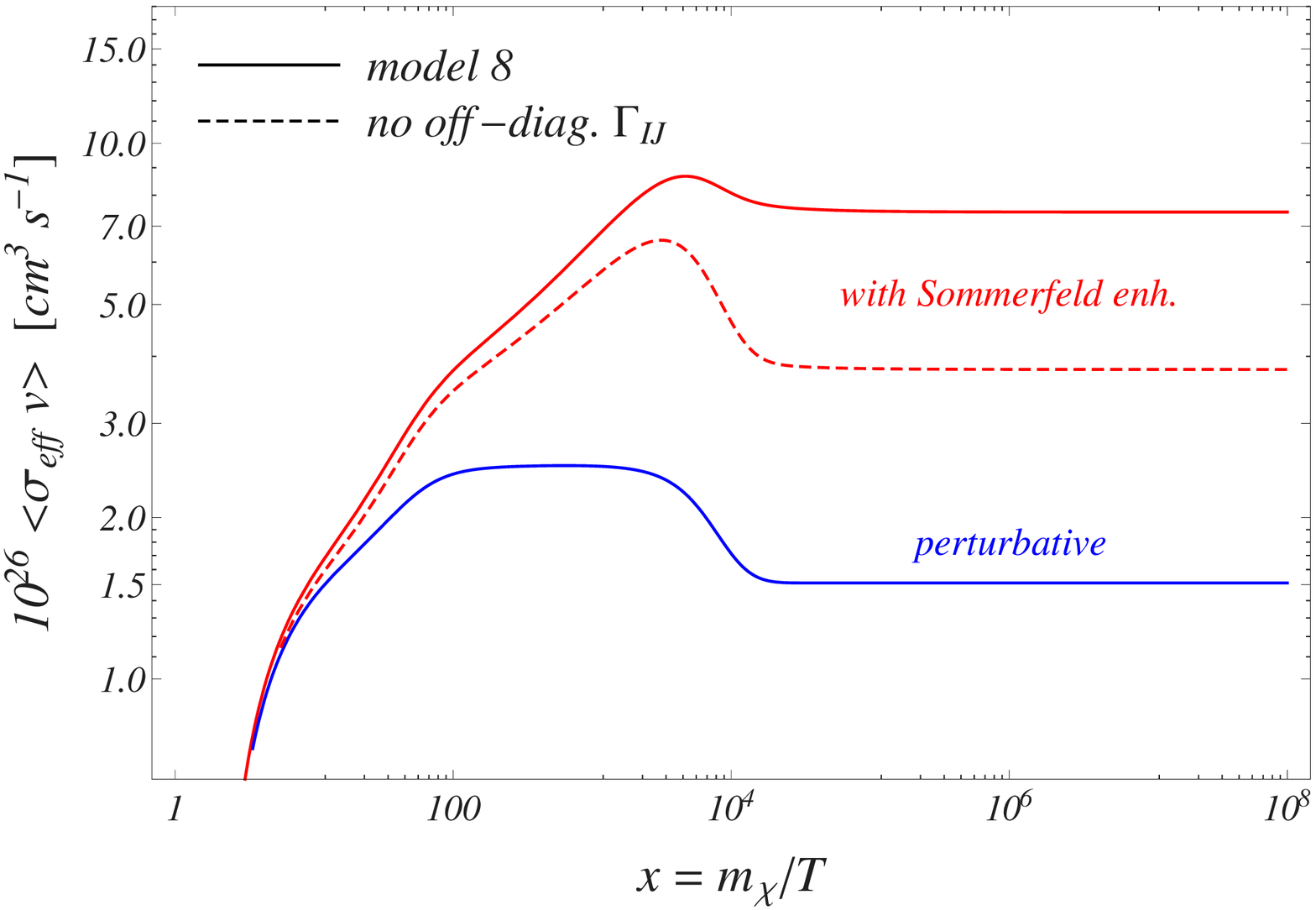}
\includegraphics[width=0.75\textwidth]{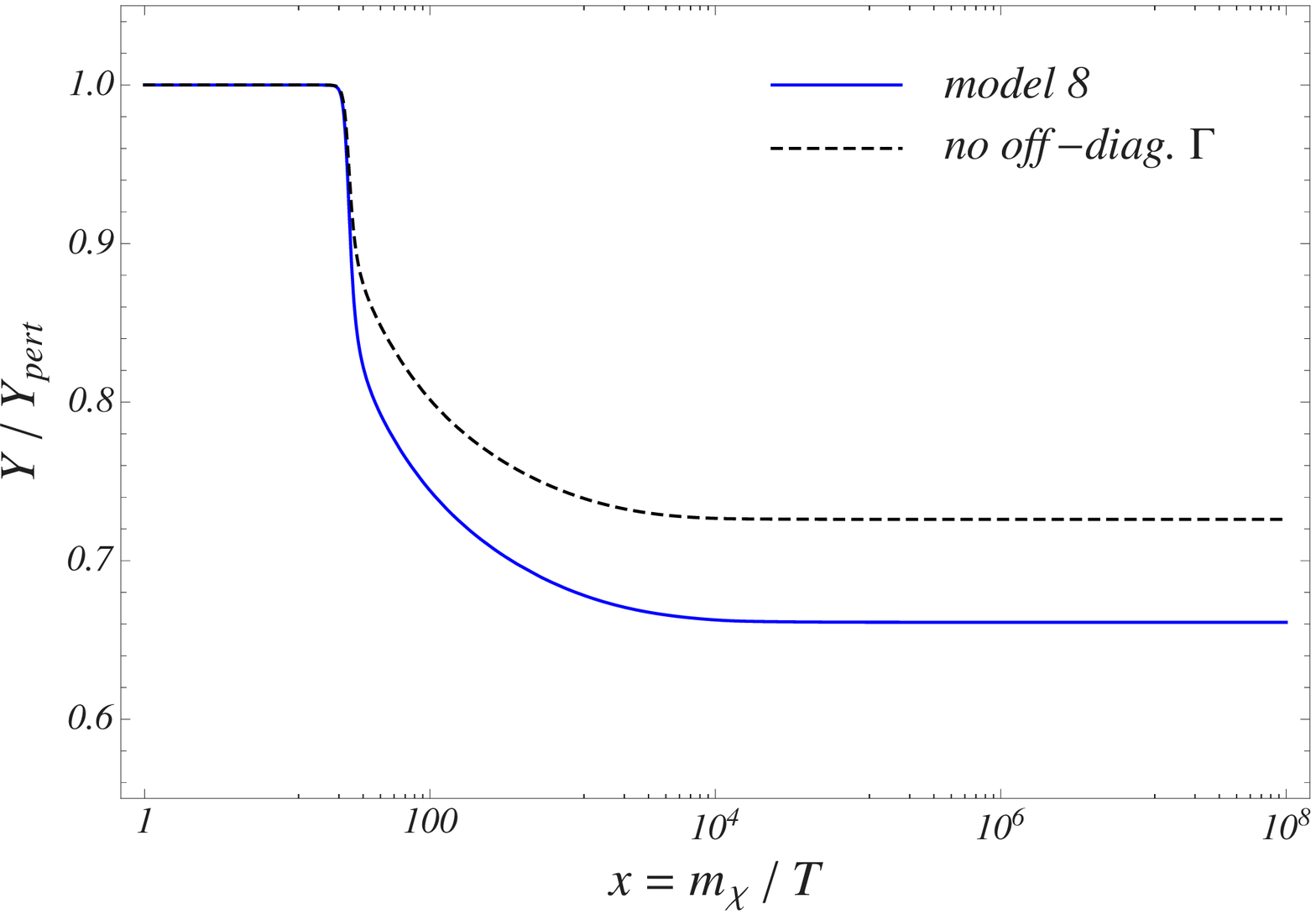}
\caption{ Upper panel: The effective thermally averaged annihilation rate
          $\langle \sigma_\text{eff} v\rangle(x)$ for trajectory model $8$.
          The two upper (red) curves show the
          $\langle \sigma_\text{eff} v\rangle (x)$ behaviour if
          Sommerfeld enhancements are taken into account
          with/without (solid/dashed) off-diagonal rates.
          The lower solid (blue) curve gives the perturbative result.
          Lower panel:
 The ratio of the yields $Y/Y_\text{pert}$ for the trajectory model
          with ID $8$ with off-diagonal rates (solid blue line) and
          without (dashed black line). 
       }
\label{fig:trajectory08_sigmaeffY}
\end{center}
\end{figure}
%---------------------------------------------------------------------------
The thermally averaged effective annihilation rates
$\langle \sigma_\text{eff} v\rangle (x)$ including (upper solid (red) line) and
neglecting (dashed red line) off-diagonal rates in the
Sommerfeld-enhanced cross sections are depicted in the upper panel of 
Fig.~\ref{fig:trajectory08_sigmaeffY}.
The corresponding perturbative result is given by the lower solid (blue) curve.
The perturbative annihilation rates of
two-particle states $\chi\chi$ heavier than the $\chi^0_1\chi^0_1$ pair are
larger than the perturbative rate of the latter, leading to a
drop in the perturbative $\langle \sigma_\text{eff} v\rangle (x)$ curve after
decoupling of the heavier co-annihilating $\chi\chi$ states.
As can be already inferred from Fig.~\ref{fig:trajectory08_sigmav}, the
effective rate including Sommerfeld enhancements turns out to be larger than
the corresponding perturbative result by factors of at most $\mathcal O(1-3)$
in the $x$ range $x = 10\ldots10^3$ relevant to the relic abundance
calculation.
These enhancements finally give rise to the behaviour of the ratio of yields
$Y/Y_\text{pert}$ shown in Fig.~\ref{fig:trajectory08_sigmaeffY}, lower panel.
Including Sommerfeld corrections on the co-annihilation rates
leads to a reduction of the relic density by $34\%$.
For this model the effect of
neglecting off-diagonal rates in the relic abundance calculation
turns out to be milder than in the wino-like $\chi^0_1$ models:
with the off-diagonal entries we get $\Omega^\text{SF}h^2 = 0.0791$
while neglecting these would lead to a value larger by $10\%$.

%-----------------------------------------------------------
\begin{table}[t]
\centering
\small
\begin{tabular}{|c|c|c|c||c|c|c||c|c|}
\hline
   neutral $\chi\chi$-states
 & $\chi^0_1\chi^0_1$
 & $\chi^+_1\chi^-_1$
 & $\chi^0_1\chi^0_2$
 & $\chi^0_2\chi^0_2$, $\chi^0_1\chi^0_3$
 & $\chi^\pm_1\chi^\mp_2$
 & $\chi^0_2\chi^0_3$
 & $\chi^0_3\chi^0_3$
 & $\chi^+_2\chi^-_2$
\\ &&&&&&&&\vspace{-1.35em}
\\
\hline
   $M_{\chi\chi}\,[$GeV$]$
 & 3688
 & 3689
 & 3745
 & 3802
 & 3803
 & 3858
 & 3915
 & 3916
\\
\hline
\hline
  charge $\pm 1$ states
 & $\chi^0_1\chi^\pm_1$
 & $\chi^0_2\chi^\pm_1$
 &
 & $\chi^0_1\chi^\pm_2$, $\chi^0_3\chi^\pm_1$
 & $\chi^0_2\chi^\pm_2$
 &
 & $\chi^0_3\chi^\pm_2$
 &
\\ &&\vspace{-1.35em}
\\
\hline
   $M_{\chi\chi}\,[$GeV$]$
 & 3689
 & 3746
 &
 & 3802
 & 3859
 &
 & 3916
 &
\\
\hline
\hline
   charge $\pm 2$ states
 & $\chi^\pm_1\chi^\pm_1$
 &
 &
 & $\chi^\pm_1\chi^\pm_2$
 &
 &
 & $\chi^\pm_2\chi^\pm_2$
 &
\\ &&&\vspace{-1.35em}
\\
\hline
   $M_{\chi\chi}\,[$GeV$]$
 & 3689
 &
 &
 & 3803
 &
 &
 & 3916
 &
\\
\hline
\end{tabular}
\\[1em]

\caption{$\chi\chi$-states and corresponding masses $M_{\chi\chi}$ in model $8$,
  ordered according to their electric charge,
  that are relevant in the calculation of the $\chi^0_1$ relic abundance
  $\Omega^\text{SF}h^2$.
  Two-particle states involving the bino-like neutralino $\chi^0_4$ are not
  shown. As their masses $M_{\chi\chi}$ lie above the scale of $20\,$TeV, they
  are irrelevant in the calculation of Sommerfeld enhancements to
  the lighter $\chi\chi$-channels and in the determination of the
  $\chi^0_1$ relic abundance. The vertical double lines separate the 
  states with masses below 3762 GeV and above 3893 GeV. 
}
\label{tab:res_spectrum-model8}
\end{table}

%-----------------------------------------------------------

It is interesting to analyse the impact on the calculated relic abundance
$\Omega^\text{SF} h^2$ when the number of channels included in the
multi-state Schr\"odinger equation is changed, or the number of heavier states
contributing to corrections from the last potential loop is varied.
Let us recall that  the results presented so far in this section
correspond to calculations where all $\chi\chi$-states with masses below
$M_\text{max}=3893\,$GeV are treated exactly in the Schr\"odinger
equation,\footnote{From the definition
$M_\text{max} = 2\,m_{\chi^0_1} + m_{\chi^0_1} v_\text{max}^2$ the quoted
value $M_\text{max}=3893\,$GeV for trajectory model $8$ is obtained by setting
$m_{\chi^0_1}=1844\,$GeV (see Tab.~\ref{tab:res_trajectory})
and $v_\text{max} = 1/3$.} 
while the remaining heavier states are included only at tree-level 
and in the last loop near the annihilation vertex in the Sommerfeld-corrected 
rates of the lighter states.
Further we have considered $\delta m^2$ corrections in the potentials for the
channels included in the Schr\"odinger equation but not in the approximate 
treatment of the heavier states (see \cite{paperIII} for details on
these corrections). In order to compare the cases
where the number of channels treated in the Schr\"odinger equation is changed,
we neglect these $\delta m^2$ corrections in the potentials throughout in the
following, so that all cases are computed with the same potential.
We calculate $\Omega^\text{SF} h^2$ for the cases of $M_\text{max}=3762\,$GeV
and $M_\text{max}=3893\,$GeV, corresponding to $v_\text{max}=0.2$ and $1/3$,
as well as for $M_\text{max}=\infty$. In the latter case all
$\chi\chi$-channels are taken into account in the Schr\"odinger equation.
To investigate the accuracy of the approximate treatment of heavier states in
the last potential loop compared to the case where these states are accounted
for exactly in the Schr\"odinger equation, we introduce the variable 
$M_\text{cut} \geq M_{\rm max}$.
$\chi\chi$-states with a mass larger than $M_\text{cut}$ are ignored 
completely. States  with mass below $M_{\rm max}$ are included in 
the Schr\"odinger equation exactly, while those with mass 
between $M_\text{max}$ and $M_\text{cut}$ are treated approximately
through the one-loop corrections in the last potential loop.
The relevant $\chi\chi$-states together with their masses are 
given in Tab.~\ref{tab:res_spectrum-model8}, from which the number of exactly
and approximately treated states in each charge sector for each of the cases
covered in Tab.~\ref{tab:res_masssplitcuts} can be read off.
The results on $\Omega^\text{SF} h^2$ that we obtain for our three
choices for $M_\text{max}$ and for $M_\text{cut}$ set 
to $M_\text{cut}=3762\,$GeV,
$3893\,$GeV and $M_\text{cut}=\infty$ are collected in
Tab.~\ref{tab:res_masssplitcuts}.

%-----------------------------------------------------------
\begin{table}[t]
\centering
\begin{tabular}{|l|c|c|c|}
\hline
 \hspace{2em} $\Omega^\text{SF} h^2$
 & $M_\text{max} = 3762\,$GeV
 & $M_\text{max} = 3893\,$GeV
 & $M_\text{max} = \infty$
\\ &&\vspace{-1.35em}
\\
\hline\hline
   $M_\text{cut} = 3762\,$GeV
 & 0.0858
 & ---
 & ---
\\
\hline
   $M_\text{cut} = 3893\,$GeV
 & 0.0817
 & 0.0816
 & ---
\\
\hline
 $M_\text{cut} = \infty$
 & 0.0804
 & 0.0801
 & 0.0801
\\
\hline
\end{tabular}
\caption{
Relic abundances $\Omega^\text{SF} h^2$ in trajectory model $8$ with a
different number of channels accounted for in the Schr\"odinger equation and
with a different number of heavy $\chi\chi$-states treated approximately in
the last potential loop.
Two-particle channels $\chi\chi$ with masses below $M_\text{max}$ are
included in the Schr\"odinger equations.
One-loop corrections of heavier $\chi\chi$-channels with masses between
$M_\text{max}$ and $M_\text{cut}$ are accounted for, while all $\chi\chi$-channels
heavier than $M_\text{cut}$ are ignored.
All results are derived neglecting $\delta m^2$ corrections in the potentials.
}
\label{tab:res_masssplitcuts}
\end{table}
%-----------------------------------------------------------
Let us first discuss the $\Omega^\text{SF} h^2$ values on the diagonal of
Tab.~\ref{tab:res_masssplitcuts}, which
display the effect of increasing the number of states in the Schr\"odinger
equation while ignoring one-loop corrections from heavier
states.
Expectedly $\Omega^\text{SF}h^2$ decreases the larger $M_\text{max}$. There are
more $\chi\chi$-channels for which Sommerfeld enhancements on their individual
annihilation cross sections are taken into account. This leads to an increase
of the thermally averaged effective rate $\langle\sigma_\text{eff} v\rangle$
entering the Boltzmann equation, which in turn decreases the relic
abundance. By increasing $M_\text{max}$ by the steps indicated in
Tab.~\ref{tab:res_masssplitcuts} the resulting $\Omega^\text{SF}h^2$ is
reduced by $5\%$ and $2\%$ respectively.
The effect on $\Omega^\text{SF} h^2$ from more channels in the Schr\"odinger
equations is rather mild as compared to the $33\%$ reduction with respect to
the tree-level relic density.\footnote{Dropping the $\delta m^2$ terms in the
  potential slightly increases the relic density for model 8 from the value
  quoted in Tab.~\ref{tab:res_trajectory}, $\Omega^\text{SF} h^2=0.0791$
  to $\Omega^\text{SF} h^2=0.0801$, which implies 
  $\Omega^\text{SF} h^2 / \Omega^\text{pert} h^2 = 0.670$.} 
The milder reduction mainly derives from the fact that the
Sommerfeld enhancement of the heavier channels' cross sections is less
pronounced than in case of the most relevant lighter channels
$\chi^0_1\chi^0_1$, $\chi^+_1\chi^-_1$ and $\chi^0_1\chi^\pm_1$.
Further, as noted previously, the heavier $\chi\chi$-channels enter the
thermally averaged rate $\langle\sigma_\text{eff} v \rangle$ with a Boltzmann
suppression factor such that their contribution is generically sub-dominant,
unless the individual rates are particularly enhanced.
The main effect that leads to the respective $5\%$ and $2\%$ change of
$\Omega^\text{SF} h^2$ comes from the slight increase of the Sommerfeld-enhanced
cross sections of the dominant light channels $\chi^0_1\chi^0_1$,
$\chi^+_1\chi^-_1$ and $\chi^0_1\chi^\pm_1$ when more states appear in the
potentials of the Schr\"odinger equations.

Let us now consider the reduction of $\Omega^\text{SF} h^2$ for fixed
$M_\text{max}$ and increasing $M_\text{cut}$. This happens because the effect
of heavier channels amounts to a positive correction to the
Sommerfeld-enhanced cross sections: the dominant potential interactions are
attractive, such that the heavier states in the last potential
loop typically give an additional positive contribution.
For instance we find a significant reduction of $\Omega^\text{SF} h^2$ by
$5\%$ from 0.858 to 0.817, when for $M_\text{max} = 3762\,$GeV the value of $M_\text{cut}$ is
increased from $3762\,$GeV to $3893\,$GeV. This indicates that the
newly added heavier states in the last loop give a large positive 
contribution to the Sommerfeld-enhanced cross sections of the 
$\chi\chi$-states in the Schr\"odinger equation.
When CPU considerations make the restriction to fewer states treated in
the Schr\"odinger equation necessary, the approximate treatment of heavy
channels should give a reasonable approximation to the case where these heavy
channels are included fully in the Schr\"odinger equation. 
This is nicely confirmed by the numbers shown in Tab.\ref{tab:res_masssplitcuts}:
when the states with mass between $3762\,$GeV and 
$3893\,$GeV are treated approximately, the reduction of $\Omega^\text{SF} 
h^2$ from 0.0858 to 0.0817 is very close to the value 
0.0816 obtained from the exact treatment of all states with mass below 
$3893\,$GeV. The same observation holds for the comparison between the 
approximate treatment of all states with masses above
$3762\,$GeV, $\Omega^\text{SF} h^2 = 0.0804$, and the exact result 
$\Omega^\text{SF} h^2 = 0.0801$. The agreement becomes even better when 
the the perturbative treatment involves only the heavier channels with mass
above $3893\,$GeV.

%-----------------------------------------------------------------------------
\section{Summary}
\label{sec:summary}

In this work we presented a detailed investigation of Sommerfeld
enhancements in the $\chi^0_1$ relic abundance calculation for several popular
models with heavy neutralino LSP in the general MSSM.
Our analysis is based on the effective field theory formalism that we
developed and described in \cite{Beneke:2012tg, Hellmann:2013jxa, paperIII}.
This framework allows us to calculate the $\chi^0_1$ relic abundance
consistently including Sommerfeld-enhanced neutralino/chargino
co-annihilation rates, taking off-diagonal rates into account and accounting for
many nearly mass degenerate co-annihilating two-particle states.
We focused on three benchmark models with wino-, higgsino- and bino-like
$\chi^0_1$ taken from \cite{Cahill-Rowley:2013gca}
as well as on a set of DarkSUSY generated spectra interpolating between the
cases of a higgsino- to a wino-like $\chi^0_1$ spectrum. With the latter set we
defined a ``higgsino-to-wino'' trajectory in the parameter space of the general
MSSM. It is worth to stress that our work allows for the first time to
investigate the Sommerfeld enhancement in neutralino/chargino
co-annihilations for a mixed wino-higgsino $\chi^0_1$.
In scenarios with wino-like $\chi^0_1$ we find a pronounced effect from Sommerfeld enhancements on the
calculated $\chi^0_1$ relic abundances, whereas for higgsino-like
$\chi^0_1$, the effect becomes milder. This is in agreement with previous
investigations in the literature in the pure-wino and pure-higgsino limits.
In general the relic abundance obtained including the Sommerfeld effect is
reduced the more the stronger the wino admixture to the $\chi^0_1$.
In addition, there are cases of particular pronounced effects related to the
existence of loosely or zero-energy bound states in the spectrum of the model.
We show that the precise value of the calculated relic density depends
on the particular details of the spectrum, such that results from a study in
the pure-wino or pure-higgsino $\chi^0_1$ scenarios do not apply directly. 
It is interesting to note that Sommerfeld enhancements in the co-annihilating
sector of a bino-like $\chi^0_1$ can affect the result on
$\Omega^\text{SF} h^2$ at the 10\% level. This is found for a 
benchmark model with bino-like $\chi^0_1$ and
slightly heavier wino-like $\chi^\pm/\chi^0$ states.
The knowledge of precise mass splittings between the co-annihilating
neutralinos and charginos is essential in the calculation of
Sommerfeld-enhanced rates and will typically require the knowledge of
spectra with a one-loop on-shell renormalised neutralino/chargino sector.

We used three pMSSM benchmark models as well as the set of models on our
``higgsino-to-wino'' trajectory in order to show the general features of
Sommerfeld-enhanced rates and their effect on the relic abundance calculation.
The results demonstrate that it will be necessary to systematically 
include the Sommerfeld effect when MSSM parameter space constraints 
on heavy neutralino dark matter from direct and indirect searches 
as well as from collider physics are combined with the requirement to 
reproduce, or at least not exceed, the observed abundance of dark matter.
A future project is the investigation of the parameter space of the general
MSSM as regards the relevance of Sommerfeld enhancements in the relic
abundance calculation. Our aim is to identify regions where the Sommerfeld
effect is not necessarily as pronounced as in the previously studied wino
limit but constitutes the dominant radiative correction. To this
end a scan of the MSSM parameter space is prepared and our findings will be
reported in future work.

%%%%%%%%%%%%%%%%%%%%%%%%%%%%%%%%%%%%%%%%%%%%%%%%%%%%%%%%%%%%%%%%%%%%%%%%%%
\subsubsection*{Acknowledgements}

This work is supported in part by the Gottfried Wilhelm Leibniz 
programme of the Deutsche Forschungsgemeinschaft (DFG) and the DFG
Sonder\-for\-schungs\-bereich/Trans\-regio~9 ``Computer\-ge\-st\"utzte
Theoreti\-sche Teilchenphysik''.
C.H. thanks the ``Deutsche Telekom Stiftung'' for its support while part of
this work was done. The work of P.~R. 
has been supported in part by the Spanish
Government and ERDF funds from the EU Commission
[Grants No. FPA2011-23778, No. CSD2007-00042
(Consolider Project CPAN)] and by Generalitat
Valenciana under Grant No. PROMETEOII/2013/007.
P.~R. also thanks
the ``Excellence Cluster Universe'' at the TU-Munich 
for its hospitality and support during completion of this work.

%%%%%%%%%%%%%%%%%%%%%%%%%%%%%%%%%%%%%%%%%%%%%%%%%%%%%%%%%%%%%%%%%%%%%%%%%%%%%%%
%\bibliography{bib-darkmatter}
%%%%%%%%%%%%%%%%%%%%%%%%%%%%%%%%%%%%%%%%%%%%%%%%%%%%%%%%%%%%%%%%%%%%%%%%%%%%%%%
\providecommand{\href}[2]{#2}\begingroup\raggedright\endgroup

\end{document}